\documentclass[useAMS,usenatbib]{mn2e}

\def\Lya {Ly$\alpha$\,}
\def\HeII {He{\sc ii}\,}
\def\Msun {\,\mathrm{M}_\odot}

\newcommand{\Zs}{$Z_{\odot}$}

\usepackage{amsmath}
\usepackage{amssymb}
\usepackage{graphicx}
\usepackage{color}
\title[Exploring the nature of CR7]{Exploring the nature of the Lyman-$\alpha$ emitter CR7}

\author[T.\ Hartwig et al.]
{\parbox{\textwidth}{
Tilman Hartwig$^{1,2}$\thanks{E-mail: hartwig@iap.fr}, Muhammad\ A. Latif$^{1,2}$, Mattis Magg$^3$, Volker Bromm$^{4}$,\\
Ralf S.\ Klessen$^{3,5}$, Simon C.\ O.\ Glover$^{3}$, Daniel J.\ Whalen$^{3,6}$, Eric W.\ Pellegrini$^{3}$,
and Marta Volonteri$^{1,2}$\\
}\\
$^{1}$Sorbonne Universit\'es, UPMC Univ Paris 06, UMR 7095, Institut d'Astrophysique de Paris, F-75014 Paris, France\\
$^{2}$CNRS, UMR 7095, Institut d'Astrophysique de Paris, F-75014, Paris, France\\
$^{3}$Universit\"at Heidelberg, Zentrum f\"ur Astronomie, Institut f\"ur Theoretische Astrophysik, Albert-Ueberle-Str.\ 2,\\
D-69120 Heidelberg, Germany\\
$^{4}$Department of Astronomy, University of Texas, Austin, Texas 78712, USA\\
$^{5}$Universit\"at Heidelberg, Interdisziplin\"ares Zentrum f\"ur wissenschaftliches Rechnen\\
$^{6}$Institute of Cosmology and Gravitation, University of Portsmouth,Portsmouth PO1 3FX, UK}

\begin{document}


\pagerange{\pageref{firstpage}--\pageref{lastpage}} \pubyear{2016}

\maketitle

\label{firstpage}

\begin{abstract}
CR7 is the brightest Lyman-$\alpha$ emitter observed at $z>6$, which shows very strong Lyman-$\alpha$ and He{\sc ii} 1640\,\AA\ line luminosities, but no metal line emission. Previous studies suggest that CR7 hosts either young primordial stars with a total stellar mass of $\sim 10^7\Msun$ or a black hole of $\gtrsim 10^6\Msun$. Here, we explore different formation scenarios for CR7 with a semianalytical model, based on the random sampling of dark matter merger trees.
We are unable to reproduce the observational constraints with a primordial stellar source, given our model assumptions, due to the short stellar lifetimes and the early metal enrichment.
Black holes that are the remnants of the first stars are either not massive enough, or reside in metal-polluted haloes, ruling out this possible explanation of CR7. Our models instead suggest that direct collapse black holes, which form in metal-free haloes exposed to large Lyman-Werner fluxes, are more likely the origin of CR7. However, this result is derived under optimistic assumptions and future observations are necessary to further constrain the nature of CR7.
\end{abstract}

\begin{keywords}
black hole physics -- stars: Population~III -- galaxies: high-redshift -- cosmology: early Universe
\end{keywords}

\section{Introduction}

The first sources of light ushered the Universe out of the cosmic dark ages, thus initiating the long history of star and galaxy formation. In the $\Lambda$CDM paradigm of hierarchical structure formation, the first generation of stars, or Population~III (Pop~III) stars, were assembled in dark matter minihaloes of a few times $10^5\Msun$.  They later merged to form more massive haloes of $\gtrsim 10^7\Msun$, which could be the birthplaces of the first galaxies \citep{bromm11}. Understanding the formation of the first galaxies is of paramount importance because they constitute the basic building blocks of present-day galaxies.

The first stars influenced subsequent structure formation by chemical, radiative and mechanical feedback. They enriched the Universe with metals, led to the formation of second-generation Population~II (Pop~II) stars, and produced energetic photons that contributed to the reionisation of the Universe. Various theoretical studies suggest that Pop~III stars were more massive than Pop~II stars because of the absence of efficient coolants in the pristine gas \citep[see the recent reviews by][]{bromm13,glover13,greif15}. Strong nebular emission lines, such as Lyman-$\alpha$ (\Lya) and those of \HeII, are expected to be present in the gas ionised by Pop~III stars \citep{tumlinson00,bromm01b,oh01,schaerer02,schaerer03}. The prime targets to detect these nebular emission lines are the first galaxies, which are expected to host both Pop~III and Pop~II stars.

Large ground based telescopes and the Hubble Space Telescope (HST) have opened a new window on high redshifts. \Lya has emerged as a powerful probe to detect distant galaxies at $z>5$. The search for high redshift galaxies has intensified over the past few years and candidate galaxies have now been detected between $z=8-11$ \citep{bouwens11,ellis13,mcl15,mcleod16}. Numerous  galaxies  have been detected  at $z>6$ using both the strong \Lya emission and the Lyman break techniques
\citep{ouchi08,ouchi09,ouchi10,vanzella11,ono12,finkelstein13,pentericci14,oesch15,matthee15}. Recently, \citet{oesch16} even claim the observation of a galaxy at $z=11.1$. Although such galaxies have been detected at these redshifts, the presence of Pop~III stars has not been observationally confirmed in any system. On the other hand, detections of high redshift quasars reveal the existence of supermassive black holes (SMBHs) of a few billion solar masses at $z>6$ \citep{fan06,mortlock11,venemans15,wu15}. Various models have been proposed to explain the formation of SMBHs which include the growth of stellar-mass black holes (BHs) as well as so-called `direct collapse' black holes (DCBHs) \citep{loeb94,bromm03,alvarez09,volonteri10,haiman13,latif13a,latif15a}. Direct observational constraints on both seed BH masses and the Pop III initial mass function (IMF) are necessary to understand the formation of the first galaxies and high redshift quasars. 

The recent discovery of strong He{\sc ii} line emission from the \Lya emitter CR7 (COSMOS redshift 7) at $z=6.6$ by \citet{sobral15} may constitute the first detection of either a Pop~III star cluster or an accreting seed BH. It is the most luminous \Lya emitter at $z > 6$ and has very strong \Lya and He{\sc ii} 1640\,\AA\ emission lines but no metal lines in the rest-frame UV. \citet{sobral15} suggest that CR7 can be explained by the composite spectra of normal metal-enriched stars and primordial stars. Deep HST imaging shows that CR7 is composed of three clumps, A, B, and C. The spectral energy distributions (SEDs) of clumps B and C are best fitted by an evolved stellar population while clump A has strong \Lya and He{\sc ii} 1640\,\AA\ lines and can be explained either by a young primordial stellar population or an accreting BH. \citet{sobral15} find that Pop~III stars with an age of a few Myr and a total stellar mass of $\sim 10^7\Msun$ with a top-heavy IMF are required to explain such strong emission lines and \citet{visbal16} demonstrate how photoionisation feedback can promote the formation of metal-free galaxies at lower redshift. The required stellar mass in Pop~III stars is degenerate with the uncertain IMF and the required stellar mass might even be higher. The possibility that CR7 hosts a BH has also been discussed in \citet{sobral15}.

\cite{pallottini15} have proposed that the strong \Lya and He{\sc ii} 1640\,\AA\ line emission in CR7 can be explained by either a $<$ 2\,Myr old Pop~III stellar cluster of $\sim 10^7\Msun$ with a top-heavy IMF or an $\sim 10^5\Msun$ BH formed at $z=7.3$.  \citet{agarwal15c}, \citet{dijkstra16}, and \citet{smidt16} have also shown that the observations can be explained by an $\sim 10^6-10^7\Msun$ BH in the progenitor haloes of CR7. But neither of their studies have shown how such a large reservoir of metal-free gas can exist at $z=6.6$ in clump A, which is required both for the formation of a young metal-free Pop~III stellar cluster or a DCBH \citep[although see][]{fop11}. 
More observations are required to find how common are sources like CR7 at such redshifts and under what conditions can they form. In addition to these further observations, a self-consistent model, which takes into account both in-situ star formation and metal enrichment, is required to better understand the assembly history and nature of CR7.

In this paper, we use a semianalytical model to examine the nature of CR7 and its assembly history. The model includes recipes for Pop~III and Pop~II star formation as well as a self-consistent treatment of metal enrichment. We consider a range of halo masses, star formation efficiencies and IMFs with a sufficient number of realisations to obtain statistically sound results. We investigate the possibility of a massive BH forming from either a Pop~III remnant or a DCBH. The results suggest that our model of Pop~III star formation, with Pop~III stars being less massive than $300\Msun$, cannot reproduce the observed \Lya and \HeII line luminosities and that a DCBH likely powers CR7.
In Section \ref{sec:obs} we review the observational constraints on CR7. We present our model in Section \ref{sec:methods} and our results in Section \ref{sec:results}. We discuss the caveats of our approach in Section \ref{sec:cav} and conclude in Section \ref{sec:conclusion}.

\section{Observational constraints}
\label{sec:obs}
CR7, the brightest Ly$\alpha$ emitter at $z > 6$, was first classified as an unreliable high redshift candidate \citep{bowler12} and as a brown dwarf candidate in the Milky Way \citep{ilbert13}. It was then found as a \Lya \,emitter candidate at $z=6.6$ by \citet{matthee15} and spectroscopically confirmed by \citet{sobral15}.
This spectroscopic follow-up revealed strong \Lya and He{\sc ii} 1640\,\AA\  lines with luminosities of $L_{\mathrm{Ly}\alpha} = (8.5 \pm 1.0) \times 10^{43}\,\mathrm{erg}\,\mathrm{s}^{-1}$ and $L_{\mathrm{HeII}} = (2.0 \pm 0.8) \times 10^{43} \mathrm{erg}\,\mathrm{s}^{-1}$. These observed luminosities imply a line ratio of \HeII /\Lya $\approx 0.23$ or a ratio of $Q(\mathrm{He}^+)/Q(\mathrm{H}) \approx 0.42$ for ionising photon emission rates. This line ratio corresponds to a very hard spectrum with an effective temperature of $T_\mathrm{eff} > 10^5$\,K, and even Pop~III stars with a mass of $1000\Msun$ only have $Q(\mathrm{He}^+)/Q(\mathrm{H}) \approx 0.1$ \citep{schaerer02}.

One possible explanation for this extreme ratio is that a fraction of the \Lya photons was not observed due to trapping in the IGM or dust absorption \citep{dijkstra06,dijkstra07,dijkstra10,zheng10,latif11b,latif11a,smith15b,smith16,matthee16}. By post-processing cosmological hydrodynamic simulations with a multiwavelength radiative transfer scheme, \citet{laursen09} and \citet{yajima14} find that the escape fraction of \Lya photons can be as low as $\sim 10\%$, with the exact value depending on specific properties such as mass, star formation, dust content, or metallicity. They also show that \Lya photons do not escape isotropically, but the flux generally depends on the direction \citep{dijkstra06}. Consequently, the observed \Lya luminosity is only a lower limit to the actual luminosity and the line ratio of \HeII /\Lya should hence be treated with caution.

In Table \ref{tab:lum}, we list line luminosities and \HeII /\Lya ratios for $1-1000 \Msun$ Pop~III stars.
\begin{table}
 \centering
 \begin{tabular}{|c|c|c|c|}
 $M_*(\Msun)$ & $L_{\mathrm{Ly}\alpha}$($\mathrm{erg}\,\mathrm{s}^{-1}$) & $L_{\mathrm{HeII}}$($\mathrm{erg}\,\mathrm{s}^{-1}$) & $L_{\mathrm{HeII}}$/$L_{\mathrm{Ly}\alpha}$\\ 
  \hline 
  \hline
  1 & $2.1 \times 10^{28}$ & $4.8 \times 10^{13}$ & $2.4 \times 10^{-15}$\\ 
  \hline 
  10 & $4.3 \times 10^{36}$ & $4.4 \times 10^{30}$ & $1.0 \times 10^{-6}$\\ 
  \hline 
  100 & $1.1 \times 10^{39}$ & $1.8 \times 10^{37}$ & $1.6 \times 10^{-2}$\\ 
  \hline 
  1000 & $1.6 \times 10^{40}$ & $3.7 \times 10^{38}$ & $2.3 \times 10^{-2}$\\ 
  \hline 
  \end{tabular} 
  \caption{Line luminosities for single, non-rotating Pop~III stars, averaged over their lifetimes with no mass loss \citep{schaerer02}. We extrapolate and interpolate linearly between the originally tabulated values. The model assumes an electron temperature, $T_\mathrm{e}$, and density, $n_\mathrm{e}$, of $30000$\,K and 100 cm$^{-3}$, which are typical values for gas around the first stars. Decreasing the electron temperature to $T_e=10000$\,K would increase the \HeII line luminosity by at most $\sim 10\%$. The luminosities are a very steep function of mass, and the line ratio of \HeII / \Lya is $\lesssim 0.02$ for the considered primordial stellar populations.}
   \label{tab:lum}
\end{table}
The line ratios of a primordial stellar population in this mass range are much smaller than the one observed for CR7. The \Lya luminosity can be reduced by trapping or absorption, but we still have to account for the very high \HeII luminosity. Put differently, we have to explain a system with $L_{\mathrm{Ly \alpha}} > 8.32 \times 10^{43}\,\mathrm{erg}\,\mathrm{s}^{-1}$ and $L_\mathrm{HeII} = 1.95 \times 10^{43} \mathrm{erg}\,\mathrm{s}^{-1}$. We will focus on the \HeII luminosity, which is more difficult to explain in the context of early structure formation, because if a primordial stellar population fulfils the \HeII constraint, it automatically satisfies the \Lya constraint.

The rate of ionising photons per stellar baryon for metal-free stars in the mass range $300 - 1000 \Msun$ is almost constant \citep{bromm01b}. In this mass range, models suggest that Pop~III stars have an effective temperature of $\sim 100$\,kK \citep{bromm01b,schaerer02} and we also see from Table \ref{tab:lum} that the hardness of the spectrum, quantified by \HeII / \Lya, does not change significantly in the mass range $100 - 1000 \Msun$. Since the binary properties and the initial rotational velocities of Pop~III stars are not well known \citep{stacy12,stacy16}, we use stellar models of single, non-rotating stars. Rotation and binary evolution might however have an influence on the evolution of massive, metal-free stars and can change their stellar lifetime, spectrum, final fate, or the remnant masses \citep{ekstroem08,deMink13}. In \citet{hartwig15b} we have investigated the effect of rapidly rotating Pop~III stars in our semi-analytical model and did not find a significant influence on the star formation or metal enrichment.

The FWHM line widths of these lines are $(266 \pm 15)\,\mathrm{km}\,\mathrm{s}^{-1}$ for \Lya and $(130 \pm 30)\,\mathrm{km}\,\mathrm{s}^{-1}$ for \HeII. Other strong radiative sources such as Wolf-Rayet stars or active galactic nuclei (AGN) generally produce broader lines with FWHM $\gtrsim 10^3\,\mathrm{km}\,\mathrm{s}^{-1}$ \citep{brinchmann08}. Recently, \citet{smidt16} demonstrate by post-processing cosmological simulations with a radiative transfer code that a massive BH of $\sim 10^7 \Msun$ accreting at $25\%$ of the Eddington limit yields a line width for \Lya that is in good agreement with the observation and a line width of $\sim 210\,\mathrm{km}\,\mathrm{s}^{-1}$ for \HeII, which is slightly above the observed value. \citet{smith16} show with a one-dimensional radiation-hydrodynamics simulation that the observed $160 \mathrm{km}\,\mathrm{s}^{-1}$ velocity offset between the \Lya and \HeII line peaks is more likely to be produced by an accreting BH than by a stellar population with an effective temperature of $10^5$\,K. Such a stellar population might ionise its environment too efficiently and can account neither for the velocity offset, nor for the spatial extension of $\sim 16\,$kpc of the \Lya emitting region. The UV slope of $\beta = -2.3 \pm 0.08$ can not be used to distinguish between different models, since both a young metal-free stellar population and an accreting BH yield a blue UV slope in this frequency range \citep{dijkstra16}.

Another striking feature of CR7 is the absence of metal lines with upper limits of, e.g., \HeII /O{\sc iii}]$1663\mathrm{\AA} > 3$ and \HeII /C{\sc iii}]$1908\mathrm{\AA} > 2.5$. This does not mean that there are no metals at all, just that the \HeII line dominates and hence normal stellar populations are excluded as the only explanation, since they would produce \HeII /O{\sc iii}]$1663\mathrm{\AA} \lesssim 0.3$, \HeII /C{\sc iii}]$1908\mathrm{\AA} \lesssim 0.3$ \citep{gutkin16}. These upper limits for the metal recombination lines also set the absolute metallicity of the gas (see Section \ref{sec:tax}). Due to the above constraints, the main source of ionising photons in CR7 has to be embedded in gas with a low metallicity.

The observed equivalent widths are EW$_{\mathrm{Ly}\alpha}>230$\,\AA \,and EW$_{\mathrm{HeII}} = (80 \pm 20)$\,\AA. The EW of \Lya is only a lower limit (since no UV continuum is detected) and has consequently no strong constraining power, because both a young stellar population and an accreting BH can yield an \Lya EW of $> 230$\,\AA\,\citep{shields93,malhotra02,schaerer03}. Also \cite{dijkstra16} do not use this observational constraint in their spectral fit since it does not yield any additional information. However, the large EW of \HeII can be used to confine the age and ambient metallicity of a potential Pop III stellar population. A comparison to the very detailed stellar evolutionary synthesis models of \citet{schaerer03} and \citet{raiter10} yields a metallicity of $<10^{-7}$ and a very recent starburst of $<1$\,Myr in order for the models to be consistent with the EW of \HeII. Unfortunately, there is not such a sophisticated model with the required parameter dependences for the EWs of an AGN spectrum. Moreover, the determination of the \HeII EW for an AGN spectrum would require the modelling of the underlying continuum. This is problematic because the AGN and stellar continuum overlap at these wavelengths and they are degenerate without knowing the respective luminosities \citep{stark15}. The detailed modelling of this problem is beyond the scope of the paper.


HST observations show that CR7 is composed of three clumps with a projected separation of $\sim 5\,\mathrm{kpc}$, with one clump hosting a young, blue stellar population and the other clumps hosting older, red populations. This might be evidence for an ongoing merger, which makes it even more important to take the merger history of this system into account. \citet{sobral15} find that the best-fitting SED model is a combination of an older $1.6 \times 10^{10}\Msun,$ 0.2 \Zs\ population with an age of $360$\,Myr and a metal-free population with a top-heavy IMF, a total stellar mass of $\sim 10^7\Msun$ and an age of a few Myr. However, a radiation source with a lifetime of at least $10-100$\,Myr is required to account for the spatial extent of the \Lya emitting region \citep{smith16}, which favours an accreting BH over a young, metal-free stellar population.

\section{Methodology}
\label{sec:methods}

We use a semianalytical model that is based on the work of \citet{hartwig15b} and enables us to efficiently test a large parameter space with high mass resolution. In this section, we present the code and introduce the techniques we use to explore the nature of CR7.

\subsection{Our basic model}

\subsubsection{Cosmological context}

According to the hierarchical scenario of structure formation, haloes merge over time to form larger structures. The distribution of halo masses as a function of redshift can be described analytically by the model of \citet{ps74}. Based on this idea, \citet{bond91} and \citet{lacey93} developed methods to construct assembly histories of individual haloes that allow the construction of dark matter merger trees. Our code is based on the merger tree algorithm by \cite{parkinson08}, which generates dark matter merger trees with arbitrary mass resolution. We use a resolution mass of $M_\mathrm{res}=2.5\times 10^5\Msun$ for the merger tree, which is sufficient to resolve all haloes for the redshifts of interest. We assume a flat $\Lambda$CDM Universe and use the \citet{planck15} cosmological parameters, most importantly the new optical depth to Thompson scattering $\tau_e = 0.066 \pm 0.016$, which is significantly lower than $\tau_e = 0.0907 \pm 0.0102$ from their previous release \citep{planck14}.

\subsubsection{Halo mass}
\label{sec:mass}
As a starting point to constructing the merger tree backwards in cosmic time, we need an approximation for the halo mass of CR7 at $z=6.6$. We use the value of $1.6 \times 10^{10}\Msun$ proposed by \citet{sobral15} as a fiducial total stellar mass for our model. This mass has been derived with SED fitting under the assumption that CR7 hosts a normal stellar population, although this might not be the case. This simplifying assumption might lead to errors of a factor of a few in the estimation of the total stellar mass. Moreover, estimates of stellar mass based on SED fitting are subject to uncertainties related to degeneracies between several stellar population parameters (star formation rate (SFR), metallicity, IMF, rotation, binarity).

To determine the halo mass for a given stellar mass we use the model by \citet{behroozi13}, who constrain SFRs and histories as a function of halo mass up to $z=8$ from observations of the stellar mass function, the cosmic SFR, and the specific SFR. For a stellar mass of $1.6 \times 10^{10}\Msun$ it yields a halo mass of
\begin{equation}
M_h = (1.2 \pm 0.2) \times 10^{12} \Msun,
\end{equation}
where the uncertainty represents the cosmological scatter. This is close to the assumed value of $M_h=10^{12}\Msun$ in \cite{agarwal15c}. We use this as a fiducial mass in our model, but also test other halo masses to check the dependence of the results on this very uncertain parameter.

\subsubsection{Star formation}
For each merger tree, we model the formation of Pop~III stars and subsequent stellar populations. For a halo to be able to form Pop~III stars, it has to fulfil four criteria. First, the halo has to be metal-free and not be polluted by in-situ star formation from progenitors or enriched externally by supernovae (we test the effect of a critical metallicity threshold in Section \ref{sec:results1}). Moreover, the halo has to be above the critical mass
\begin{equation}
 M_\mathrm{crit} = 3 \times 10^5 \left( \frac{T_\mathrm{crit}}{2.2 \times 10^3\,\mathrm{K}} \right) ^{3/2} \left( \frac{1+z}{10} \right) ^{-3/2} M_\odot
 \label{eq:mcrit}
\end{equation}
to enable the primordial gas to cool efficiently and trigger star formation, where we have normalised the critical temperature to the value found in cosmological smoothed particle hydrodynamics simulations of \citet{hummel12}. Furthermore, mergers dynamically heat the gas and can delay Pop~III star formation. Hence, the mass growth rate of a certain halo has to be below
\begin{equation}
 \frac{\mathrm{d} M}{\mathrm{d} z} \lesssim 3.3 \times 10^6 \Msun \left( \frac{M}{10^6 M_\odot} \right)^{3.9}
 \label{eq:mdyn}
\end{equation}
to enable primordial star formation \citep{yoshida03}, or otherwise star formation is delayed. Finally, Lyman-Werner (LW) photons can also delay or even prevent the collapse of primordial gas by photodissociating H$_2$. Following \citet{machacek01}, the fraction of the total gas mass that cools and collapses under the influence of a LW background is
\begin{equation}
 f_\mathrm{LW}= 0.06 \ln \left( \frac{M_\mathrm{halo} / M_\odot}{1.25 \times 10^5 + 8.7 \times 10^5 F_\mathrm{LW} ^{0.47}} \right),
 \label{eq:fLW}
\end{equation}
where $F_\mathrm{LW}$ is the LW flux in units of $10^{-21}\mathrm{erg}\ \mathrm{s}^{-1} \mathrm{cm}^{-2} \mathrm{Hz}^{-1}$. Primordial star formation is suppressed if $f_\mathrm{LW}$ falls below zero.

Once we have identified a Pop~III star-forming halo, we assign individual stars to it whose masses are randomly sampled from a logarithmically flat IMF between $M_\mathrm{min}$ and $M_\mathrm{max}$. The shape of the IMF is motivated by numerical simulations that predict a relatively flat mass distribution that is dominated by high-mass stars \citep{clark11a,greif11,susa14,hirano14,hirano15} and the total stellar mass in a halo with virial mass $M_\mathrm{vir}$ is
\begin{equation}
M_* = \eta f_\mathrm{LW} \frac{\Omega _b}{\Omega _m} M_\mathrm{vir},
\label{eq:SFE}
\end{equation}
where $\eta$ is the star formation efficiency (SFE) parameter for primordial star formation.

We use spectra tabulated by \citet{schaerer02,schaerer03} to determine the number of ionizing and LW photons and the \Lya and \HeII line luminosities for each Pop~III star individually as a function of its mass. Lifetimes for the stars are also taken from \citet{schaerer02,schaerer03}. The associated line luminosities for Pop~III stars are listed in Table \ref{tab:lum}. We adopt luminosities averaged over the stellar lifetime, which are smaller than the zero-age main sequence luminosities by a factor of a few. This might be relevant for a recent burst in Pop~III stars but does not change our final conclusions.

We model the formation of subsequent generations of stars (Pop~II) from the observed cosmic star formation history at high redshift and account for their chemical and radiative feedback (ionising and LW photons). We take cosmic SFRs from \citet{behroozi15}, extrapolate them to $z>15$ and set them to zero for $z>30$.

\subsubsection{Escape fractions}

For the escape fraction of ionising radiation we use $f_{\rm esc,III}=0.5$ for Pop~III star-forming haloes and $f_{\rm esc}=0.1$ for later generations of stars in more massive haloes. Generally, the escape fraction depends on the halo mass, redshift and the stellar physics \citep{paardekooper13,trebitsch15}, but we use these average values for our simplified model, which are in good agreement with previous studies \citep{johnson09,finkelstein12,wise14,renaissance16}. A higher escape fraction of ionising radiation would lead to a lower SFE, which only weakly affects the final results as we show in Section \ref{sec:results2}.

Escape fractions for LW photons are taken from \citet{schauer15}, who find that they are a function of both halo and Pop~III star mass. The escape fraction is dominated by the most massive star in each halo, which first ionises the surrounding gas. Hence, we use the escape fraction values in the far field approximation for the most massive star in each halo.

\subsubsection{Metal enrichment and critical metallicity}
\label{sec:metal}

After its lifetime and depending on its mass, a Pop~III star explodes in a supernova (SN) and pollutes its environment with metals. We use metal yields from \citet{heger10} and assume a Sedov-Taylor expansion of the enriched volume. By summing the volumes of these SN remnants in all haloes, we calculate the time-dependent fraction of the total volume that is already polluted. Once a new halo forms, we check statistically whether it is already metal-enriched or still pristine. A key assumption of this model is that the SN remnant is still in the Sedov-Taylor phase when it breaks out of its host halo. This is valid for minihaloes, as the H{\sc ii}  region expansion has already cleared most of the gas from the halo \citep{wan04,ket04,abs06,awb07}. For later generations of stars in more massive haloes, we assume that they do not contribute to the pollution of the IGM by metals \citep[although see][]{mf99}.

In principle, a halo can host several Pop~III SNe and their metal-enriched remnants can overlap. Since the expansion of the enriched gas is mainly dominated by the most massive (and consequently first) SN to go off \citep{ritter15}, we only account for the expansion of the SN of the most massive progenitor per halo. This approximation is valid, as long as there are not too many highly energetic SNe in one halo, which is generally not the case in our models.

We are not only interested in whether a halo is polluted by metals, but also in the metallicity of the polluted gas. This information enables us to allow Pop~III star formation not only in pristine gas but also below a certain critical metallicity $Z_\mathrm{crit}$. Due to the lack of spatial information in the merger tree, we use the approximation that all metals are deposited in the outer shell of the expanding SN. Hence, we calculate a metal surface density ($\Sigma _m (t)$) for each SN explosion that decreases as the SN expands, and we construct a time dependent probability distribution for these surface densities. When we find that a halo is polluted with metals, we randomly draw a $\Sigma _m (t)$ from the probability distribution and determine the mass of the metals with which the newly formed halo is polluted from $M_m = \Sigma _m (t) \pi R_\mathrm{vir}^2$, where $R_\mathrm{vir}$ is its virial radius. Assuming that these metals mix homogeneously with the gas, we calculate its metallicity as follows:
\begin{equation}
\label{eq:Z}
Z= \frac{M_m}{0.02\,M_\mathrm{vir}\,\Omega _b / \Omega _m},
\end{equation}
which yields $10^{-6}\lesssim Z \lesssim 10^{-2}$ for external metal enrichment by SNe. The accretion and inflows of pristine gas on to the halos are taken into account self-consistently due to the smooth accretion of gas below the resolution limit of the merger tree. Note that from this point forward we state all metallicities in units of the solar metallicity, $Z_\odot$.

\subsubsection{SFEs based on merger history}

As an alternative to a constant SFE, we include another recipe for star formation based on the merger history of the haloes. It has been shown that mergers can enhance star formation because of tidal torques, which allow the efficient transport of gas to the centre of the galaxies where the dense gas can cool and form stars \citep[e.g.,][]{sanders96,cox08}. This is further supported by observations, which show a negative correlation between star formation indicators and the projected distance of galaxies \citep{barton00,lambas03,smith07}. To account for this effect, we use the model of \citet{cox08} to determine the SFR. \citet{cox08} study the effect of the galaxy mass ratio on merger-driven starbursts with numerical simulations for typical galaxies in the local Universe. For disc galaxies with halo masses in the range $(0.5 - 11.6)\times 10^{11}\Msun$, the burst efficiency is best described by a fit of the form
\begin{equation}
\eta _\mathrm{burst} = \epsilon _\mathrm{1:1} \left( \frac{M_\mathrm{sat}}{M_\mathrm{primary}} \right) ^{\alpha},
\label{eq:bursteff}
\end{equation}
where $\epsilon _\mathrm{1:1}$ is the burst efficiency for equal mass mergers, $M_\mathrm{sat}$ is the mass of the satellite and $M_\mathrm{primary}$ the mass of the primary. The stellar mass per halo is then
\begin{equation}
M_* = \eta _\mathrm{burst} \frac{\Omega _b}{\Omega _m} M_\mathrm{vir}.
\label{eq:SFE2}
\end{equation}
For the local Universe and specified mass range, they propose values of $\epsilon _\mathrm{1:1}=0.55$ and $\alpha=0.69$. Applying these values to higher redshifts and less massive haloes leads to drastic overestimates of SFRs and premature reionisation ($\tau_e = 0.287$). $\epsilon _\mathrm{1:1}$ depends more strongly on the absolute mass of the galaxies than $\alpha$, so we treat it as a free parameter to match the reionisation history of the Universe and keep $\alpha$ constant.

\subsubsection{Sampling the halo mass function and calibrating SFEs}

To determine the star formation efficiencies $\eta$ and $\epsilon _\mathrm{1:1}$ we calibrate them against $\tau_e$, which is a measure of the integrated ionisation history of the Universe. The optical depth is very sensitive to the number of ionising photons in the early Universe and can therefore be used to calibrate the SFE. To do so, we sample the halo mass function at $z=6.6$ from $10^8-10^{13}\Msun$ and weight the number of ionising photons from a given halo by the number density of haloes of this mass at that redshift. This yields a cosmologically representative sample, since haloes with masses below $10^8\Msun$ at $z=6.6$ hardly produce any ionising photons and haloes above $10^{13}\Msun$ are very rare at this redshift. For more details and a thorough comparison to analytical models of the halo mass function see \citet{magg16}.

For 11 equally distributed halo masses in this range we generate 100 merger trees, which yield a statistically representative number of realisations. We then calculate $\tau_e$ from the ionisation history of these haloes. We chose the SFE accordingly, to match the observed value of $\tau_e=0.066$. Reionisation is mainly driven by Pop~II stars, but we also need the contribution by primordial stars. We do not account for other sources in the total ionising budget in the early Universe, such as quasars \citep[e.g.,][]{volonteri09,mh15} or high-mass X-ray binaries, and the uncertainty in $\tau_e$ might yield different SFEs. However, we demonstrate in Section \ref{sec:results2} that a different SFE has no influence on metal enrichment or our final conclusions.

For this implementation of the halo mass function and 100 randomly generated merger trees we probe a cosmologically representative volume of $\sim 10^6\,\mathrm{Mpc}^{-3}$. This is statistically sufficient for most of our purposes, but might be too small to probe certain rare scenarios of SMBH seed formation (see Section \ref{sec:DCBH}).

\subsection{Models of Pop~III star formation}

We model the star formation history of CR7, focusing on the primordial stellar component, with our semianalytical code. To investigate the possibility of having $10^7\Msun$ of Pop~III stars at $z=6.6$ and the corresponding high \HeII line luminosity and EW we test several models of primordial star formation, which are summarised in Table \ref{tab:models}.
\begin{table}
 \centering
 \begin{tabular}{|c|c|c|c|c|}
  label & $M_\mathrm{min}$ & $M_\mathrm{max}$ & SFE & $Z_\mathrm{crit}$ \\ 
  \hline 
  \hline
  fiducial & $3\Msun$ & $300\Msun$ & Eq. \ref{eq:SFE}, $\eta=0.14$ & 0\\ 
  \hline
  1$\rightarrow$100 & $1\Msun$ & $100\Msun$ & Eq. \ref{eq:SFE}, $\eta=0.20$ & 0\\ 
  \hline 
  $Z_\mathrm{crit}$ & $3\Msun$ & $300\Msun$ & Eq. \ref{eq:SFE}, $\eta=0.12$ & $10^{-3.5}$\\ 
  \hline 
  merger & $3\Msun$ & $300\Msun$ & Eq. \ref{eq:SFE2}, $\epsilon _\mathrm{1:1}=4 \times 10^{-3}$ & 0\\
  \hline 
  \end{tabular} 
  \caption{Overview of the four models we use for primordial star formation. The fiducial model assumes an IMF from $3-300\Msun$ and only pristine gas can form Pop~III stars. In the second model, we change the IMF to slightly lower masses from $1-100\Msun$. The third model allows Pop~III star formation up to a metallicity of $Z_\mathrm{crit}=10^{-3.5}$. In the last model, we determine the stellar mass per halo based on the merger history. In all the models, we calibrate the SFE to reproduce the optical depth to Thomson scattering.}
\label{tab:models}
\end{table}
For the `fiducial' model we assume a logarithmically flat IMF from $M_\mathrm{min}=3\Msun$ to $M_\mathrm{max}=300\Msun$ and an SFE $\eta = 0.14$, which yields an optical depth $\tau_e = 0.067$. The mass range of the IMF is consistent with recent simulations \citep{greif11,susa14,hirano14,hirano15} and it covers all possible stellar remnants relevant to our model. In a second model, which we label `1$\rightarrow$100', we assume a lower-mass IMF from $M_\mathrm{min}=1\Msun$ to $M_\mathrm{max}=100\Msun$, motivated by recent simulations of primordial star formation \citep{clark11a,stacy12,dopcke13,hartwig15a,stacy16}, which manifest disc fragmentation and hence lower-mass Pop~III stars. This model should illustrate the effects of a different mass range of primordial stars, since the actual values are not well constrained. In this second model, we adopt $\eta = 0.20$ which results in $\tau_e = 0.066$.

The main difference between primordial and later generations of star formation is the ability of the gas to cool efficiently. Metal lines can cool the gas to lower temperatures than cooling by molecular hydrogen, which is the most efficient coolant in primordial gas. Hence, the Jeans mass in metal enriched gas is smaller than in primordial gas and the cloud consequently fragments into more and smaller clumps, which then collapse to form protostars. So far we assumed that Pop~III stars form from pristine gas with a top-heavy IMF, but several studies show that even with trace amount of metals a top-heavy IMF is possible \citep{bromm01a,schneider02,schneider12,frebel07,dopcke13,safranek14}. 

Consequently, in the third scenario, named `$Z_\mathrm{crit}$', we assume that Pop~III stars form with a top-heavy IMF out of metal enriched gas with a metallicity of $Z<Z_\mathrm{crit}$ with $Z_\mathrm{crit} = 10^{-3.5}$. Although dust cooling can yield a lower $Z_\mathrm{crit}$ \citep{schneider12,dopcke13}, we adopt this value as a conservative upper limit. We use a flat IMF from $M_\mathrm{min}=3\Msun$ to $M_\mathrm{min}=300\Msun$. The effective temperature is lower for a metal enriched star than for a primordial star of the same mass so its spectrum is softer \citep{bromm01b}. The LW and \Lya luminosities are only weakly affected but the \HeII line luminosity is generally smaller for $0< Z \lesssim Z_\mathrm{crit}$ compared to the metal-free case \citep{cojazzi00,schaerer03}. On average, it is smaller by a factor of $\sim 10$, but the exact value depends on the treatment of stellar winds and metallicity. Due to this uncertainty, we will still use the line luminosities for the metal-free case but keep in mind that this yields a strict upper limit for the \HeII line. We also verified that the final results are insensitive to the choice of $Z_\mathrm{crit}$. This is in agreement with \citet{latif15c}, who show that the fraction of haloes that are enriched up to a certain metallicity is only a weak function of the actual metallicity for $10^{-6} < Z < 10^{-4}$. An SFE $\eta=0.12$ yields an optical depth $\tau_e = 0.068$.

In the fourth model, which we label `merger', we couple the star formation to the merger history based on equation \ref{eq:SFE2} with $\epsilon _\mathrm{1:1}=4 \times 10^{-3}$. For mergers with a mass ratio below $0.2$, we set $\epsilon _\mathrm{1:1}=10^{-5}$, which allows about one star to form per halo. The IMF extends from $M_\mathrm{min}=3\Msun$ to $M_\mathrm{min}=300\Msun$ and only pristine gas can form Pop~III stars. We use radiative feedback in this model only to check if a minihalo can collapse in a given LW background, but we do not use this value to determine the final mass in Pop~III stars as we do in the other models. We obtain an optical depth of $\tau_e = 0.067$ with this model.

One should keep in mind that our understanding of primordial star formation is still quite incomplete and uncertain due to the lack of any direct observations. We try to overcome this uncertainty by implementing different scenarios of Pop~III star formation that cover the most likely theories about the formation of the first stars. We have also tested other parameters, such as a primordial IMF extending to masses above $300 \Msun$ (see Section \ref{sec:results1}), but since these results deviate even further from the observational constraints of CR7, we do not explicitly discuss them here.

\subsection{Pop~III remnant black hole}

We also investigate the possibility that CR7 hosts a massive BH. \citet{pallottini15} show that a BH with an initial mass of $\sim 10^5\Msun$ embedded in a halo of total gas mass $10^7\Msun$ can account for the observed line luminosities about $100$\,Myr after formation. This result was derived by coupling a 1D radiation-hydrodynamic code \citep{pacucci15a} to the spectral synthesis code {\sc cloudy} \citep{ferland13} as described in detail in \citet{pacucci15b}. After $\sim 100$\,Myr the gas in the halo is depleted and the \HeII luminosity decreases. Generally it is possible to obtain the necessary fluxes at later times if the reservoir of metal-poor gas is large enough. First, we explore the formation scenario in which massive BHs grow from Pop~III stellar remnants \citep{madau01,haiman01,volonteri03a,whalen12}.

The final fate of a star depends mainly on its mass. Pop~III stars with masses of $25 \Msun \leq M_* \leq 140 \Msun$ or with $M_* > 260\Msun$ collapse to BHs, although the exact ranges depend on rotation and magnetic field strengths \citep{karlsson13}. $140-260\Msun$ Pop~III stars are completely disrupted in pair instability SNe that leave no remnant behind.  Primordial stars that directly collapse to a BH do not pollute their host halo with metals. This is an important characteristic of primordial stars, which facilitates having an accreting BH in a metal-free environment.

We trace the formation of Pop~III remnant BHs in the assembly history of CR7 and merge the BHs if the mass ratio of a merger of two galaxies is $>0.1$ \citep{taffoni03,volonteri03a,wassenhove14}. For smaller mass ratios, we only follow the growth of the more massive BH. The rate at which Pop~III remnant BHs grow by accretion is a subject of ongoing debate and depends on the gas supply, the depth of the gravitational potential well of the halo, the merger history, and the radiative feedback \citep{milos09a,milos09b,pm11,pm12,whalen12,pm13,pacucci15b}. Existing simulations have indicated that accretion on to stellar-mass Pop~III remnants may be substantially suppressed, due to radiation-hydrodynamical feedback \citep{johnson07,milos09a,milos09b,jeon12,jeon14}. Hence, we do not take gas accretion into account and note that the derived values of the BH masses are a strict lower limit. We discuss the issue of gas accretion on to Pop~III remnant BHs in more detail in Section \ref{sec:cav}.

\subsection{Direct collapse black hole}
\label{sec:DCBH}
As a second scenario of massive BH formation we study the direct collapse model, in which a $\sim 10^5\Msun$ seed BH is formed as a consequence of rapid isothermal collapse \citep{bromm03,begelman06,latif13a,shlosman15}. In this picture, a sufficiently strong LW background photodissociates molecular hydrogen, which otherwise triggers Pop~III star formation in 10$^5$ - 10$^7 \Msun$ haloes. Without cooling by H$_2$, a minihalo can grow until it reaches a virial temperature of $\sim 10^4$\,K, when atomic hydrogen cooling becomes efficient. If the LW flux is above a critical value, $J_\mathrm{crit}$, the gas collapses isothermally at very high central infall rates that form a supermassive star, which then collapses to a BH. We check if a halo is metal-free and if its virial temperature is above $10^4$\,K, which implies a minimum halo mass of \citep{glover13}
\begin{equation}
M_\mathrm{atom} = 4.5 \times 10^7 \Msun \left( \frac{z+1}{11} \right)^{-3/2}.
\end{equation}
To compute the LW flux, we adopt the model of \citet{dijkstra14} and \citet{habouzit15}. We assume that the flux is provided by one nearby star-forming halo, for which the LW luminosity is
\begin{equation}
L_\mathrm{LW} = 10^{47}h \bar{\nu}\,\mathrm{s}^{-1} \frac{M_*}{\mathrm{M}_\odot} \left(1+ \frac{t_6}{4} \right)^{-3/2} \exp \left( - \frac{t_6}{300} \right),
\end{equation}
where $h\bar{\nu} = 2\times 10^{-11}\,\mathrm{erg}$ is the mean energy of a LW photon, $M_*$ is the stellar mass of the halo, and $t_6$ is the time in Myr after the initial starburst. Here, we assume that $5\%$ of the gas in the halo turns into stars, which is in agreement with the model by \cite{behroozi15} for the redshifts and halo masses of interest. The collapse time of the atomic cooling halo is approximately 10\,Myr \citep{visbal14}, which is equal to the minimum time for which a LW flux $>J_\mathrm{crit}$ is required. Since the production of LW photons decreases with time after the initial starburst, we use this time as a minimum requirement to produce sufficient LW photons ($t_6=10$). Hence, the distance up to which a star-forming halo of mass $M_h$ can provide a flux $\geq J_\mathrm{LW}$ is
\begin{equation}
r_\mathrm{rad} = 48\,\mathrm{kpc} \left( \frac{M_h}{10^{11}\Msun} \right)^{1/2} \left( \frac{J_\mathrm{LW}}{100} \right)^{-1/2}.
\end{equation}
We use the general convention to express the LW flux in units of $10^{-21}\,\mathrm{erg}\,\mathrm{s}^{-1}\,\mathrm{cm}^{-2}\,\mathrm{Hz}^{-1}\,\mathrm{sr}^{-1}$.

\begin{figure}
\centering
\includegraphics[width=0.47\textwidth]{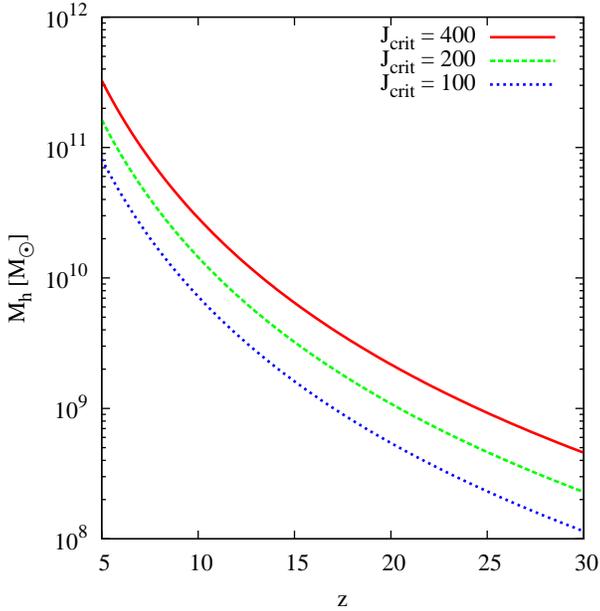} 
\caption{Minimum halo mass required to produce $J_\mathrm{crit}$ without polluting its neighbour halo with metals, plotted as a function of redshift. Depending on the redshift and the required LW flux, a halo of $\sim 10^9-10^{11}\Msun$ is needed to suppress H$_2$ formation in its neighbour and produce a DCBH.}
\label{fig:J21}
\end{figure}

For a given star-forming halo, we calculate the pollution radius, which provides a minimum distance between this halo and the atomic cooling halo. Assuming a Sedov-Taylor expansion of the metal enriched galactic winds into the surrounding gas, which has a density of $\Delta = 60$ times the mean density of the intergalactic medium \citep{dijkstra14}, the radius of metal enrichment can be expressed as
\begin{equation}
r_\mathrm{metal} = 22\,\mathrm{kpc} \left( \frac{M_h}{10^{11}\Msun} \right)^{1/5} \left( \frac{1+z}{11} \right)^{-6/5}.
\end{equation}
The necessary requirement to form a DCBH is $r_\mathrm{metal} < r_\mathrm{rad}$, which translates into a minimum mass of
\begin{equation}
M_h > 7.2 \times 10^9\Msun \left( \frac{1+z}{11} \right)^{-4} \left( \frac{J_\mathrm{LW}}{100} \right)^{5/3}
\label{eq:mass21}
\end{equation}
for a nearby star-forming halo to provide the LW flux $J_\mathrm{LW}$. This minimum mass is plotted in Figure \ref{fig:J21} as a function of redshift for a variety of fluxes.
For the considered critical fluxes, halo masses of $\sim 10^9-10^{11}\Msun$ are required to provide sufficient photodissociating radiation. The haloes have to be more massive at lower redshifts because the ambient density of the haloes decreases with time, which in turn increases the radius of metal pollution. The critical flux required for isothermal collapse proposed in the literature spans several orders of magnitude and depends on the detailed physics of collapse and the radiation spectrum \citep{sugimura14,glover15a,glover15b,agarwal15b,latif15a,hartwig15c}. Assuming a Pop~II starburst in a $\sim 10^{9}\Msun$ halo about 10\,Myr ago and a distance to this halo of $\gtrsim 10$\, kpc, we expect $J_\mathrm{crit} = 100-700$ \citep{agarwal15b}. For $J_\mathrm{crit} \gtrsim 600$ we find no DCBHs in our model, because of the limited cosmological volume that we can simulate. Hence, we vary $J_\mathrm{crit}$ from 100--400, and discuss our choice of $J_\mathrm{crit}$ in more detail in Section \ref{sec:cav}.

To identify formation sites of DCBHs we first find a metal-free halo with a mass $\geq M_\mathrm{atom}$ and its most massive neighbour in the merger tree. These are two haloes at the present time step that merge in the next time step. This condition of an incipient merger guarantees a small spatial distance between the two haloes. If, for a given $J_\mathrm{crit}$, the mass of the nearby halo fulfils equation \ref{eq:mass21}, we assume that it provides a sufficient LW flux without polluting the atomic cooling halo and that a DCBH forms.

\citet{agarwal15c} propose that CR7 hosts a DCBH. In $\sim20\%$ of their merger tree realisations, which represent the mass assembly histories of CR7, a DCBH may form. The formation redshift of the DCBH is $z\sim20$, which is limited by two factors: at higher redshift, the LW flux, which they calculate from \citet{agarwal15b}, is not high enough. At lower redshift the formation site of the DCBH is polluted by metals, where they assume that metals are ejected at a constant wind speed of $100\,\mathrm{km}\,\mathrm{s}^{-1}$. For a seed mass of $2 \times 10^4\Msun$, accretion at $40\%$ of the Eddington rate, and an escape fraction of $f_\mathrm{esc}=0.16$ for \Lya photons, they show that a DCBH is able to reproduce the observed line luminosities of \Lya and \HeII. Hence, if the BH is able to accrete metal-free gas for long enough, formation at higher redshifts is also possible. Following \citet{pallottini15} and \citet{agarwal15c}, we assume that a DCBH formed before $z=7.3$ can account for the observed line luminosities if it accretes low-metallicity gas at $z=6.6$. Motivated by \citet{latif13c} and \citet{ferrara14}, we assume that DCBHs form with an initial seed mass of $10^4-10^5\Msun$ in our model. These masses, however, are plausibly upper limits, as the strength of the LW flux we assumed is lower than assumed in those papers. \citet{latif15b}, for instance,  find that lower LW fluxes result in lower mass concentrations in the precursors of DCBHs when $J_{21}<1000$.

\subsection{Determination of the metal tax}
\label{sec:tax}
\begin{figure}
\centering

\includegraphics[angle=-90,width=0.47\textwidth]{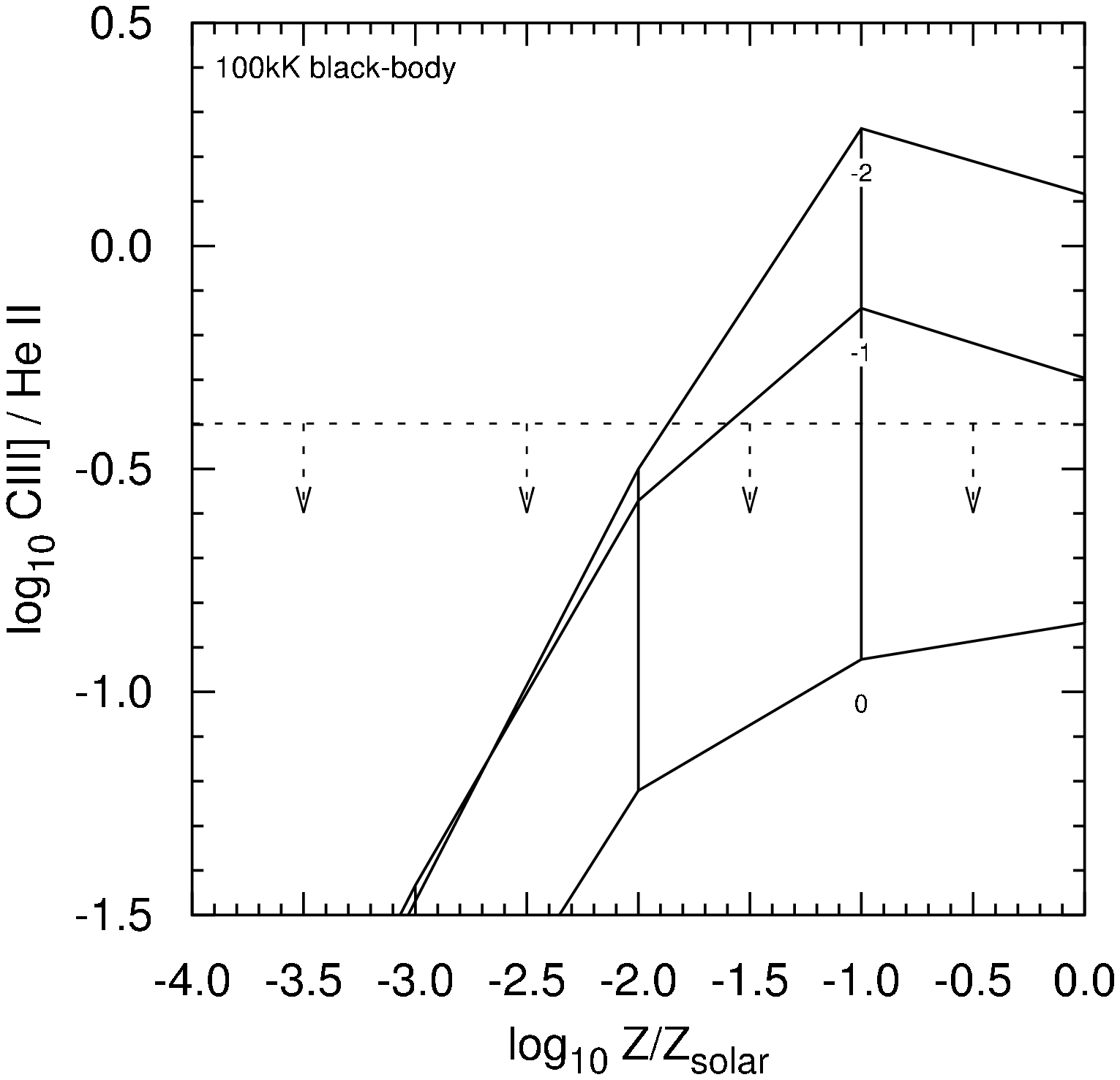}
\includegraphics[angle=-90,width=0.47\textwidth]{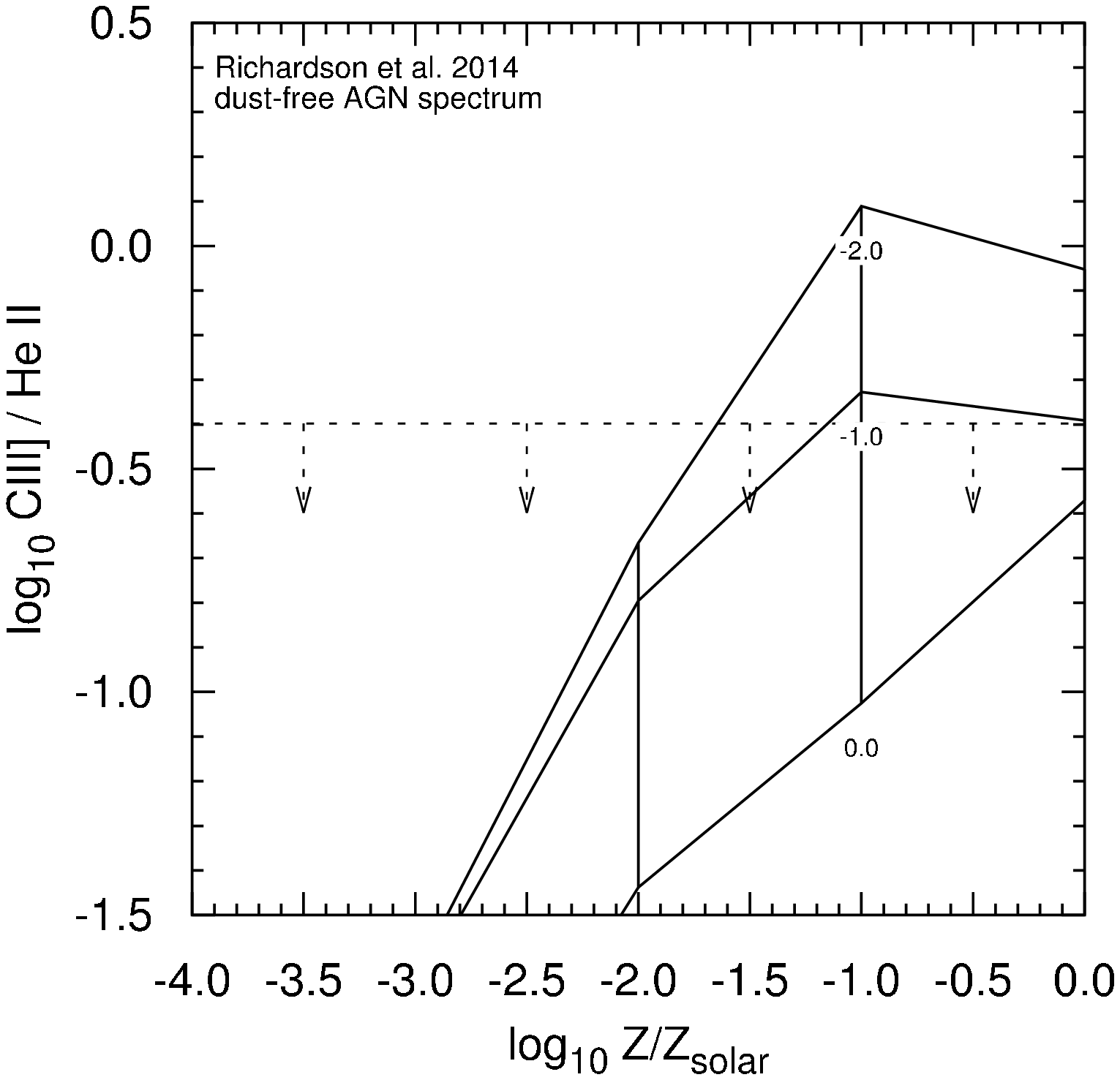}

\caption{Line ratio as a function of the gas metallicity for a 100kK blackbody (top) and an AGN SED (bottom). The dashed line indicates the upper observational limit of C{\sc iii}]$1908\mathrm{\AA}$/\HeII $< 0.4$ and the three different lines represent different ionisation parameters of the gas. For metallicities below $\sim 10^{-2}$, all models are in agreement with the observation.}
\label{fig:Zlimit}
\end{figure}

A crucial parameter to understand the nature of CR7 is the `metal tax', respectively the maximum tolerable metallicity that does not violate the observational limits on the metal lines. We use the photoionisation code {\sc cloudy}, version 13.03 \citep{ferland13} to calculate the metal line ratios as a function of the gas metallicity and dust abundance, approximating that the dust abundance and the individual elemental abundances all scale linearly with the total metallicity. We assume an ambient gas density of $10\,\mathrm{cm}^{-3}$ and have checked that the results are only weakly affected by a higher density of $100\,\mathrm{cm}^{-3}$. For the Pop~III stellar population, we adopt a 100\,kK blackbody spectra, since most of the contribution to the recombination lines of interest might come from such massive, hot stars. To model the emission from the accretion disc of a BH, possibly residing in CR7, we assume an AGN SED based on \citet{richardson14}. This model is tuned to achieve reasonable agreement along the AGN sequence by matching the \HeII /H$\beta$ ratios of observed AGN. The accretion disc of this AGN can be modelled as a multicolour blackbody with a maximum temperature of $T_\mathrm{max}\approx 7 \times 10^5$\,K. We also tested a different SED to verify that the choice of the AGN model only has minor influence on the line ratio, in which we are interested in. The resulting line ratios as a function of metallicity can be seen in Figure \ref{fig:Zlimit} for different ionisation parameters ($U$) and for the two spectral models.
The C{\sc iii}]$1908\mathrm{\AA}$/\HeII ratio is a strong function of the gas metallicity and the ionization parameter. For a high ionisation parameter of $\log U = 0$, even solar metallicity gas is in agreement with the observed line ratios. For lower ionisation parameters of $-2 \lesssim \log U \lesssim -1$, as we expect to find for high redshift AGNs \citep{nagao06,feltre15}, lower metallicities are required. Given the uncertainties in the model, it is safe to assume
\begin{equation}
Z_\mathrm{limit}=10^{-2}
\end{equation}
as an upper limit of the metallicity in clump A for both scenarios.

\begin{figure}
\centering
\includegraphics[angle=-90,width=0.47\textwidth]{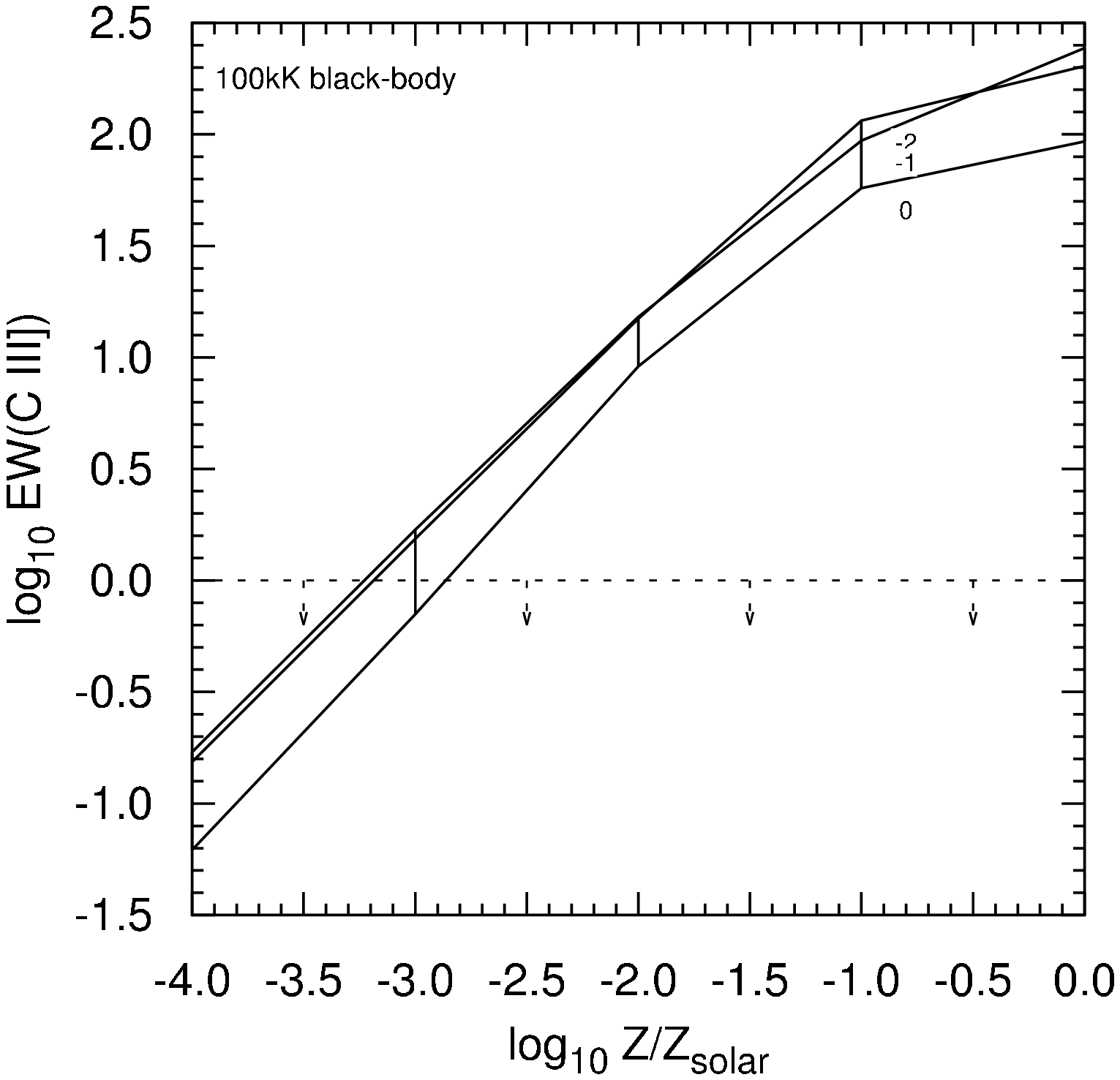} 
\hspace{5mm}
\includegraphics[angle=-90,width=0.47\textwidth]{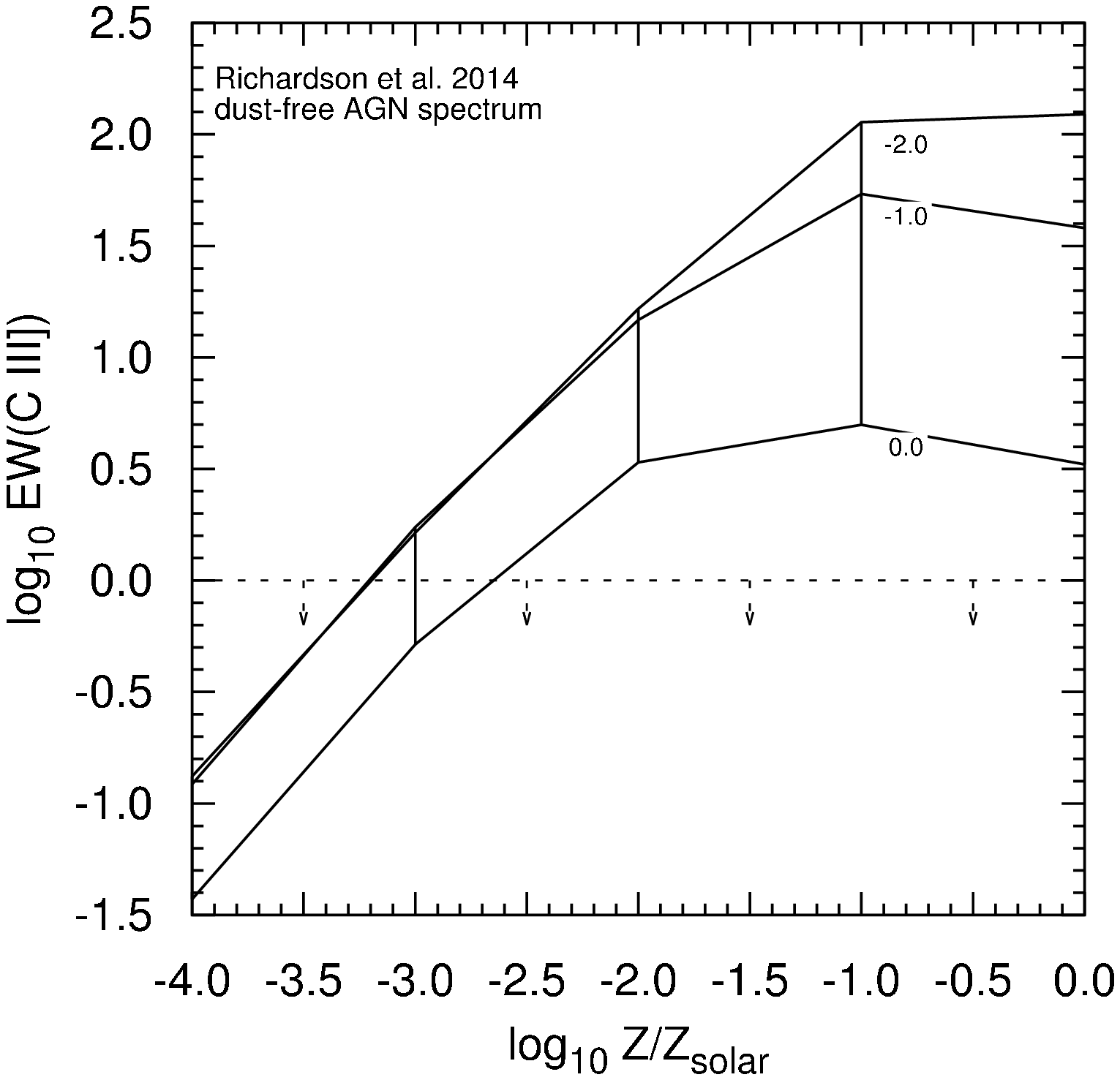} 
\caption{Modelled EW of C{\sc iii}] as a function of the gas metallicity for a stellar population (top) or a AGN spectrum (bottom). Given the spectral resolution, the EW should be below $\lesssim 1\,\mathrm{\AA}$, which sets an upper limit to the gas metallicity of $\sim 10^{-3}$.}
\label{fig:CIII}
\end{figure}

A different approach is to further investigate the non-detection of the C{\sc iii}] doublet at $\sim 1908\mathrm{\AA}$. For the spectral resolution of $0.4\,\AA$ \citep{sobral15}, we expect the EW of C{\sc iii}] to be below $\lesssim 1\,\AA$, since it should otherwise be detected as an emission line in the spectrum. This yields an additional constraint on the gas metallicity  as can be seen in Figure \ref{fig:CIII}.
From the Cloudy model at different ionisation parameters we can constrain that only gas metallicities of $\lesssim 10^{-3}$ are consistent with the non-detection of the C{\sc iii}] doublet. Although this is even more constraining than our previous approach based on the line ratios, we use the conservative value of $Z_\mathrm{limit}=10^{-2}$, due to the difficulties in modelling the EW with our simple model. We check the influence of a possibly lower metal tax in the final discussion.

\section{Results}
\label{sec:results}

In this section, we present the main results of our analysis and investigate whether CR7 can be explained with our models of Pop~III star formation or by an accreting BH. For each set of parameters, we create 100 independent merger tree realisations and average the derived quantities over the different realisations (if not stated otherwise). This yields a cosmologically and statistically representative sample with statistical scatter of $<10\%$.

\subsection{Cosmologically representative models of primordial star formation}
\label{sec:results1}

We sample the halo mass function from $10^8-10^{13}\Msun$ and weight the contribution to the number of ionising photons by the number density of those haloes at $z=6.6$. This enables us to reproduce the observed value of the optical depth.

\begin{figure}
\centering
\includegraphics[width=0.47\textwidth]{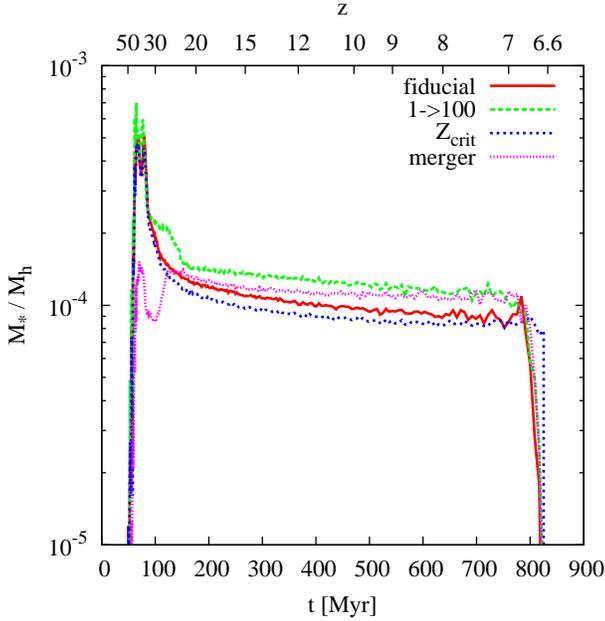} 
\caption{Effective SFE for Pop~III stars as a function of cosmic time for the four models. After a small early peak, the SFE is approximately constant around $10^{-4}$.}
\label{fig:zeta}
\end{figure}

The total mass of Pop~III stars per halo depends on either the LW background (equation \ref{eq:SFE}) or on the merger history (equation \ref{eq:SFE2}). To compare these recipes for star formation, we define the effective SFE as $\eta _\mathrm{eff}=M_*/M_h$, which is shown in Figure \ref{fig:zeta} as a function of cosmic time.
Typical values for $\eta_\mathrm{eff}$ are $\sim 10^{-4}$ for all the models. Minihaloes have masses on the order of $\sim 10^6 \Msun$, and for $\eta_\mathrm{eff} \sim 10^{-4}$ we form a total stellar mass of $\sim 100\Msun$ on average per Pop~III star-forming minihalo. The random sampling of the IMF leads to statistical variance from halo to halo and there are haloes with multiple Pop~III stars and masses $>100\Msun$. Indeed, the majority of primordial stars in our model form in multiples of 2-6 stars, as predicted by simulations of early star formation \citep{clark11a,greif11,stacy12,dopcke13,hartwig15a,stacy16}.

\begin{figure}
\centering
\includegraphics[width=0.47\textwidth]{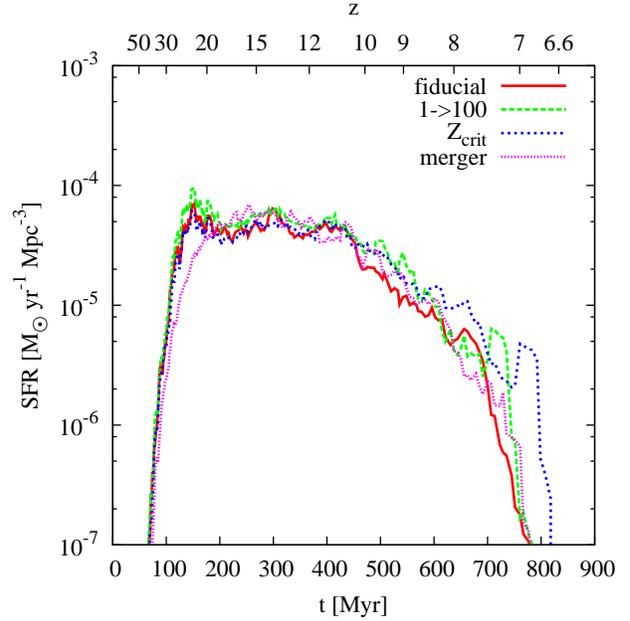} 
\caption{Mean cosmic Pop~III SFR for all four models per comoving volume. The primordial SFR peaks around redshift 20, and decreases at $z<12$.}
\label{fig:zSFR}
\end{figure}

The cosmic mean Pop~III star formation density as a function of time for all four models is shown in Figure \ref{fig:zSFR}.
It is roughly the same for all the models because they are constructed to satisfy the latest constraints on $\tau_e$. The $Z_\mathrm{crit}$ model's values are slightly higher just shortly above $z=6.6$ because Pop~III stars also form in low-metallicity gas and hence also at later times. Consequently, this model is less affected by metal pollution at lower redshifts. The derived cosmic star formation densities are in agreement with those of \citet{visbal15}. They show that the recent constraints on $\tau_e$ by the \citet{planck15} limit the mean cosmic star formation density of primordial stars to $\lesssim 10^{-4} \Msun\,\mathrm{yr}^{-1}\,\mathrm{Mpc}^{-3}$. The SFR density at $z\sim 6.6$ might not be cosmologically representative due to ongoing Pop~III star formation at $z\lesssim 6.6$, which is not captured by our method.

\begin{figure}
\centering
\includegraphics[width=0.47\textwidth]{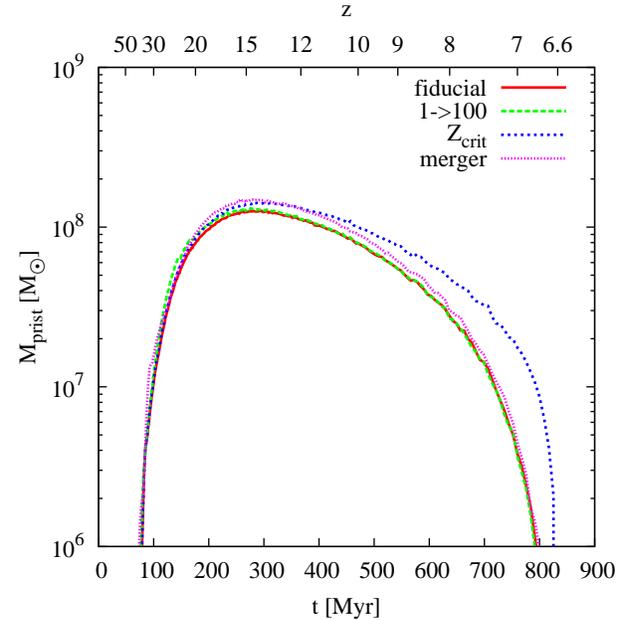} 
\caption{Mass of metal-free or metal-poor gas as a function of time for all four models. Here, we assume that the final halo has a mass $M_h=1.2 \times 10^{12}\Msun$ at $z=6.6$ and we only account for gas in its resolved progenitor haloes, which explains the rise at early times. This plot illustrates the maximum available amount of gas to form Pop~III stars, but even in the most promising $Z_\mathrm{crit}$ model the mass of gas available for Pop~III star formation is limited to $\lesssim 10^8\Msun$.}
\label{fig:tMprist}
\end{figure}

The fact that we have more gas available to form primordial stars in the $Z_\mathrm{crit}$ model can be seen in the plots of metal-poor gas mass as a function of time in Figure \ref{fig:tMprist}.
At a given redshift, this is the sum of the metal-free gas in all resolved haloes in the merger tree. Hence, it is a measure of the maximum available mass to form Pop~III stars, assuming an SFE of $100\%$. For all the models this mass remains below $\sim 10^8 \Msun$. In the model in which we allow primordial stars to also form in metal-enriched gas at $Z<Z_\mathrm{crit}$, we have more gas to form Pop~III stars at lower redshifts and still $10^7 \Msun$ of low-metallicity gas just above $z=6.6$. These values are derived for $M_h = 1.2 \times 10^{12}\Msun$.

\begin{figure}
\centering
\includegraphics[width=0.45\textwidth]{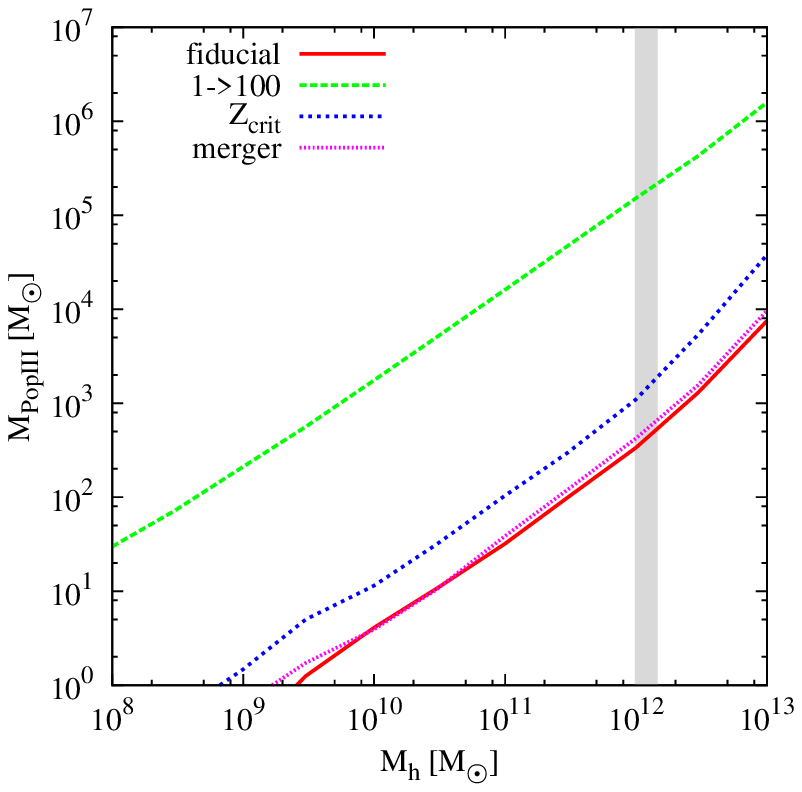} 
\includegraphics[width=0.45\textwidth]{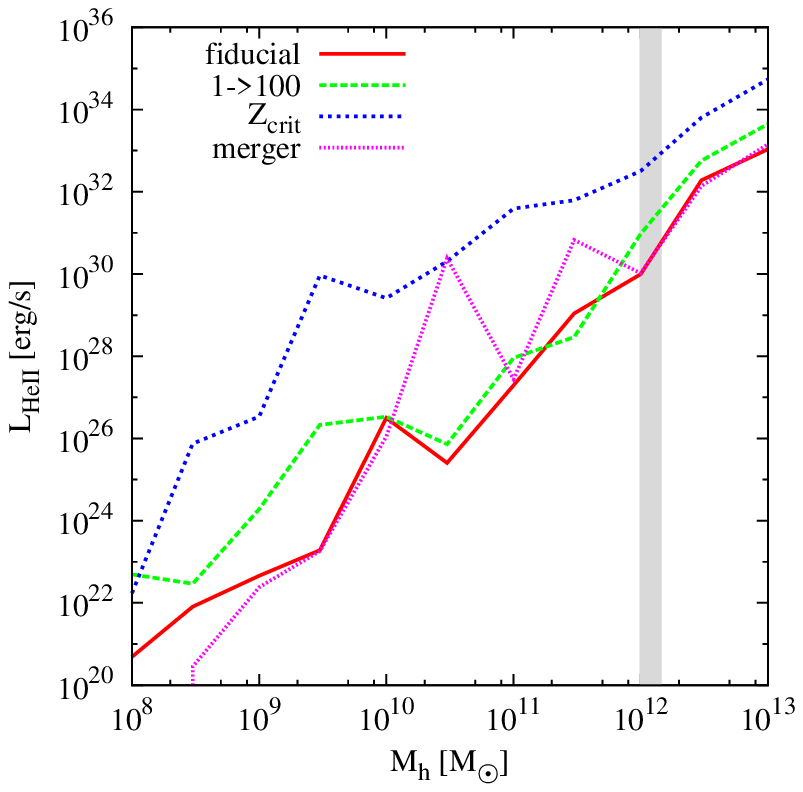} 
\caption{Stellar masses in Pop~III stars (top) and corresponding \HeII luminosities (bottom) at $z=6.6$ for a variety of halo masses. Here, we illustrate the entire range of simulated halo masses, and the most likely mass range for CR7 is shaded in grey. The model with a lower mass IMF produces a significantly higher stellar mass in Pop~III stars, mainly because of the longer lifetimes of these stars. In the expected mass range of CR7, the stellar primordial mass is limited to $10^2-10^5\Msun$, depending on the model. The $Z_\mathrm{crit}$ model produces the highest \HeII luminosity, because this model is less affected by metal enrichment at lower redshifts and can hence form Pop~III stars out to later times. The \HeII  luminosity is significantly lower than the observed value of $1.95 \times 10^{43} \mathrm{erg}\,\mathrm{s}^{-1}$.}
\label{fig:M}
\end{figure}

The stellar mass in Pop~III stars and \HeII line luminosities are shown in Figure \ref{fig:M}.
For the lower mass IMF there are $10^5\Msun$ of Pop~III stars in CR7, whereas the other models yield values of $10^2-10^3\Msun$. This is in agreement with the results of \citet{xu16}, who find in their cosmological simulation $\lesssim 10^3\Msun$ of Pop~III stars in halos at $z=7.6$. The mass of pristine gas drops steeply before this redshift and only less massive stars with longer lifetimes can survive to be present in CR7. The \HeII luminosity is a steep function of the stellar mass, and massive stars are favoured to reproduce the observations. The $Z_\mathrm{crit}$ model yields larger masses of low-metallicity gas down to smaller redshifts and consequently allows primordial star formation at later times. Hence, more massive stars can also survive to contribute to the \HeII luminosity at $z=6.6$. The $Z_\mathrm{crit}$ model produces the highest \HeII luminosities, which are still more than 10 orders of magnitude below the observed value of $1.95 \times 10^{43} \mathrm{erg}\,\mathrm{s}^{-1}$. Even for a higher halo mass, which might be possible within the uncertainties of the determination of the stellar and halo mass of CR7 (see Section \ref{sec:mass}), the final luminosities are too small. We note that the corresponding \Lya luminosity for Pop~III stars in a $\sim 10^{12}\Msun$ halo is of the order of $10^{36}-10^{37}\,\mathrm{erg}\,\mathrm{s}^{-1}$, which is about 7 orders of magnitude below the observed value.

\begin{figure}
\centering
\includegraphics[width=0.47\textwidth]{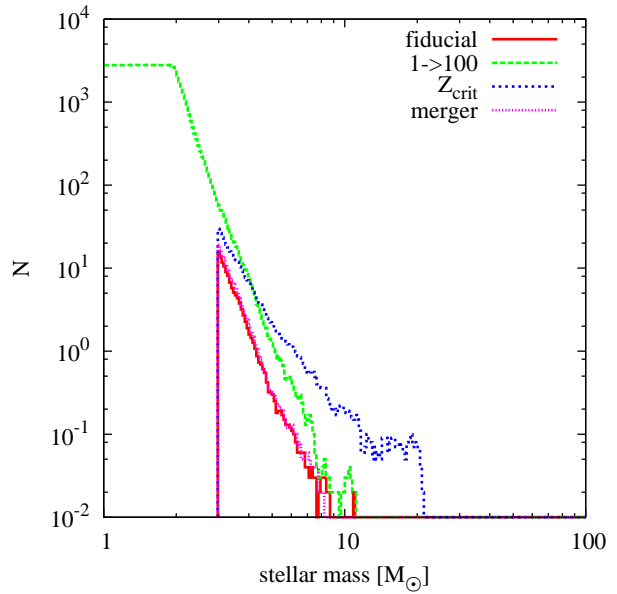} 
\caption{Number of stars per logarithmic mass bin that are present in the final $1.2 \times 10^{12}\Msun$ halo at $z=6.6$, summed over 100 realisations. The more massive the stars, the shorter their lifetimes and the lower the probability that they survive from their time of formation until the final redshift. The model with the less massive IMF produces many low-mass stars, which survive down to $z =$ 6.6 if they are less than $\sim 2\Msun$. In the $Z_\mathrm{crit}$ model primordial stars form up to just above $z=6.6$, so more massive survivors are present in the final halo.}
\label{fig:IMF}
\end{figure}

What mostly limits the \HeII luminosity is the mass of Pop~III stars that survive until $z=6.6$. To show this effect, we plot the stellar mass distribution of primordial stars in Figure \ref{fig:IMF}.
The more massive the stars, the shorter the lifetimes and the smaller the probability that they survive long enough to be present down to $z=6.6$. For $M<2\Msun$ this plot represents the IMF, since the lifetimes of these stars are long enough for them to survive until $z=6.6$. For higher masses we see that the $Z_\mathrm{crit}$ model is the most likely one to also contain stars that are $\sim 10 \Msun$ since it has the largest amount of gas available for star formation down to lower redshifts. But even these stars are not massive enough to contribute significantly to the \HeII luminosity, due to the steep dependence of the \HeII luminosity on the stellar mass (see also Table \ref{tab:lum}). We also test more extreme models for the Pop~III IMF with a mass range from $M_\mathrm{min}=10\Msun$ to $M_\mathrm{max}=1000\Msun$ and find significantly fewer primordial stars at $z=6.6$ and also a smaller \HeII line luminosity than in the fiducial model. For $M_\mathrm{min} \gtrsim 50\Msun$ there are no Pop~III stars at all that might contribute to the \HeII luminosity at $z=6.6$, because such massive stars explode within a few Myr as SNe.

\begin{figure}
\centering
\includegraphics[width=0.47\textwidth]{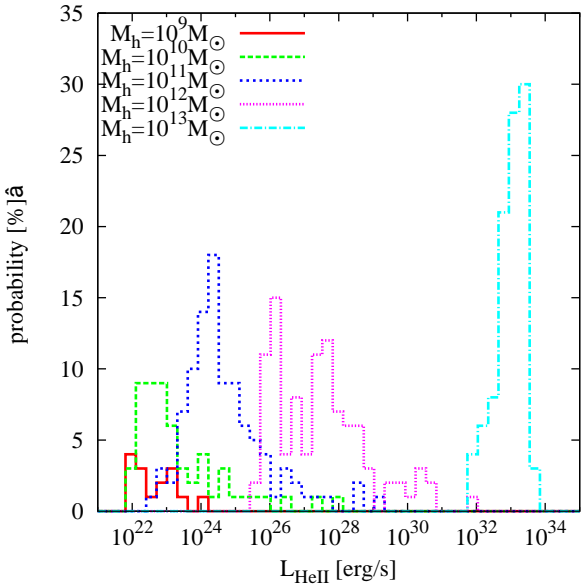}
\caption{Probability distribution function for the expected \HeII line luminosities in the merger model for a variety of final halo masses. Due to the random sampling of merger histories, the scatter in the final \HeII luminosity in this model is broader compared to the other models. However, even the scatter and the associated probabilities for large values of $L_\mathrm{HeII}$ cannot account for the missing 10 orders of magnitude between model and observation.}
\label{fig:histo}
\end{figure}

Coupling star formation to the merger history of the haloes induces a higher scatter in the stellar mass per halo. Hence, the merger model leads to a broader distribution of \HeII luminosities at $z=6.6$. We show the probability distribution function of the luminosities at this redshift in Figure \ref{fig:histo}.
The expected \HeII luminosities for a halo mass of $M_h=10^{12}\Msun$ span more than six orders of magnitude with a maximum value of $\sim10^{32}\,\mathrm{erg}\,\mathrm{s}^{-1}$. The recipe for star formation in mergers (equation \ref{eq:SFE2}) is only an extrapolation to lower-mass haloes and higher redshifts. But even in the most optimistic case, in which all the pristine gas turns into Pop~III stars during a major merger of two $\sim 10^6 \Msun$ minihaloes, $\sim 100$ such mergers are required just above $z=6.6$ to account for $10^7\Msun$ of Pop~III stars at that redshift. It is therefore difficult to explain the \HeII emission in CR7 with our models for primordial star formation.

\subsection{Alternative scenarios of primordial star formation}

\label{sec:results2}
\begin{figure}
\centering
\includegraphics[width=0.47\textwidth]{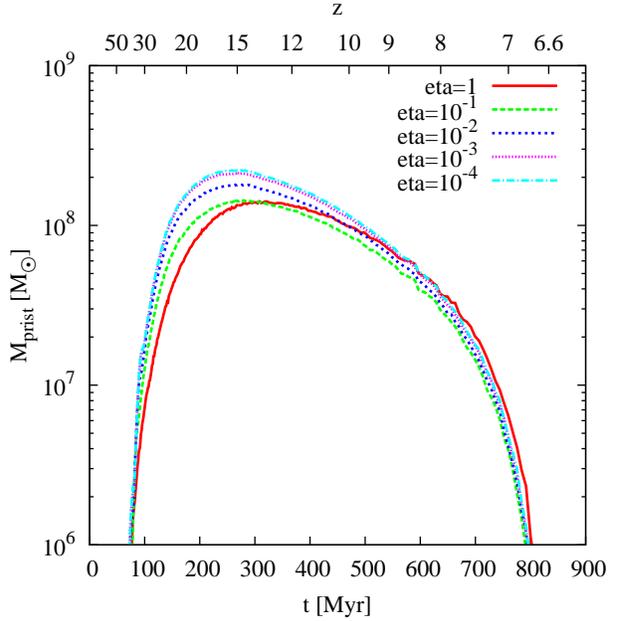}
\caption{Mass of pristine gas in a $1.2 \times 10^{12}\Msun$ halo as a function of time for a range of SFE parameters. The probability that a halo pollutes its environment with metals is not a strong function of the SFE, but only depends on the probability that there is at least one highly-energetic SN in the halo.}
\label{fig:eta}
\end{figure}

The host halo of CR7 corresponds to a rare $\sim 3\sigma$ peak in the cosmological density field at $z = 6.6$. Such rare halos have a comoving number density of only $\sim 10^{-7}$\,Mpc$^{-3}$ at this redshift, and it is possible that star formation within these rare objects might proceed differently than in the average galaxy at this redshift.
So far, we have assumed that CR7 is cosmologically ``representative" and our star formation recipes reproduce the reionisation history of the Universe. Since CR7 cannot be explained in this way, we now drop the constraint of the optical depth, which enables us to vary the SFE of primordial star formation. In other words, we no longer require the mean Pop III SFE in the progenitors of CR7 to be the same as the global mean required to produce the right Thomson scattering optical depth, but instead treat it as a free parameter. We show the effect of changing the SFE in Figure \ref{fig:eta}.
Even if we change the SFE by 4 orders of magnitude, the mass of zero-metallicity gas is only weakly affected. In none of the models is this mass sufficient to explain the observational signatures of CR7, and the \HeII line luminosity is limited to $\lesssim 3 \times 10^{32}$ erg s$^{-1}$. To understand this rather weak dependence of pristine gas mass and final luminosity on the SFE, we investigate the two limiting cases for the metal enrichment model. As described in Section \ref{sec:metal}, we only account for the SN expansion of the most massive star per halo, which could either be a core-collapse SN with an explosion energy of $E_0 = 1.2 \times 10^{51}\,\mathrm{erg}$, or an pair-instability supernova (PI SN) with $E_0 = 10^{52}-10^{53}\,\mathrm{erg}$, depending on the mass of the star. For a very low value of $\eta$, we form only one star per halo and the probability that this star explodes is $\sim 45\%$ for a logarithmically flat IMF from $3-300\Msun$. For a very high value of $\eta$, we form many Pop~III stars per halo and have a correspondingly high probability to obtain a star that explodes as a highly-energetic PI SN. The radius of the metal enriched volume is approximately proportional to $E_0^{1/5}$ so it is only weakly affected by SNe with different explosion energies. This explains the small variations in the mass of pristine gas between the models with different SFEs. Having several SNe going off in one halo might break the conservative assumption of our metal enrichment model that the pollution of metals is dominated by the most massive SN. Considering multiple SNe per minihalo would lead to even less pristine gas at lower redshifts.

We have assumed that CR7 is one halo at $z=6.6$, although we clearly see three distinct clumps. By construction, the final halo is polluted by metals because its progenitors were enriched by SNe or external metal enrichment. Consequently, there is no Pop~III star formation at $z=6.6$. In an alternative scenario, we now assume that CR7 is an ongoing merger and that the three clumps will merge to one halo in 100\,Myr ($z_0=6.0$) or 200\,Myr ($z_0=5.6$). Based on \cite{behroozi13}, we estimate the halo masses of halo A and C to be $M_A=(3 \pm 0.6) \times 10^{10}\Msun$ and $M_C=(6 \pm 1.2) \times 10^{11}\Msun$. The merger time of these haloes with the more massive halo B are $\gtrsim 50$\,Myr at this redshift \citep{boylan08}. For $z_0 \lesssim 5.6$, the masses of the third and fourth most massive halo tend to be approximately equal, which does not match the constellation of CR7, where we only observe three clumps (see also Section \ref{sec:comp}). This limits the possible range to $5.6 \lesssim z_0 \lesssim 6.0$.

\begin{figure}
\centering
\includegraphics[width=0.47\textwidth]{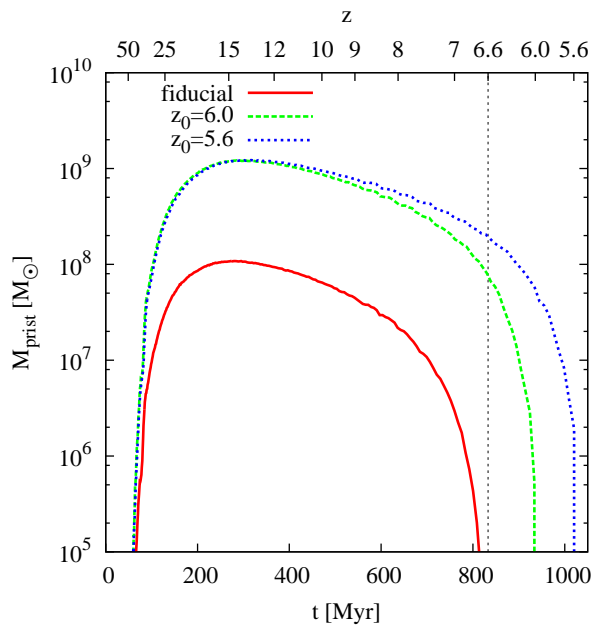}
\caption{Mass of pristine gas as a function of cosmic time for models that assume a final mass of $1.2 \times 10^{12}\Msun$ but several final redshifts, for which the merger trees are created backwards in time. The gas mass available at this redshift is much higher for those models, but we would still need an SFE of 10\% and to overcome other problems (see text).}
\label{fig:zz}
\end{figure}

These redshifts are the starting points for constructing merger trees backwards in time so that we can determine the primordial stellar mass and corresponding \HeII luminosity at $z=6.6$. The primordial gas mass is shown as a function of time in Figure \ref{fig:zz}.
The primordial gas mass at $z=6.6$ is much higher for the haloes whose merger tree extends down to redshifts $z_0<6.6$. The two additional models yield $\sim 10^8\Msun$ of pristine gas at $z=6.6$. The corresponding total masses of Pop~III stars are $2 \times 10^4 \Msun$ and $1 \times 10^5\Msun$, which are 2-3 orders of magnitude higher than in the fiducial model but still not sufficient to account for the observational constraints. The \HeII luminosities are $7 \times 10^{39}\,\mathrm{erg}\,\mathrm{s}^{-1}$ and $3 \times 10^{40}\,\mathrm{erg}\,\mathrm{s}^{-1}$ for the $z_0=6.0$ and $5.6$ models, respectively. If Pop~III stars form instantaneously out of the pristine gas with $10\%$ efficiency just above $z=6.6$, the mass of pristine gas in these models would be sufficient to explain observations of CR7. However, there is no plausible mechanism that could trigger an instantaneous starburst of this intensity, which is required to explain both the luminosity and the EW of the \HeII emission.

Although the models with $z_0 < 6.6$ could in principle explain CR7, there are two shortcomings. First, the primordial stars are distributed over all the CR7 progenitors. In our model, we just add up all the Pop~III stars at $z=6.6$ although they should be confined to clump A. This additional constraint would limit the Pop~III stars (and their total mass) to just this one clump (see Section \ref{sec:comp}). Another effect becomes important at lower redshifts that we have not considered. After the reionisation of the Universe, photoionisation heating counteracts the cooling in minihaloes and might prevent their collapse \citep{abel99,pawlik09}, which further limits the number of Pop~III stars that can form at lower redshifts.

\subsection{Pop~III remnant black holes}

\begin{figure}
\centering
\includegraphics[width=0.47\textwidth]{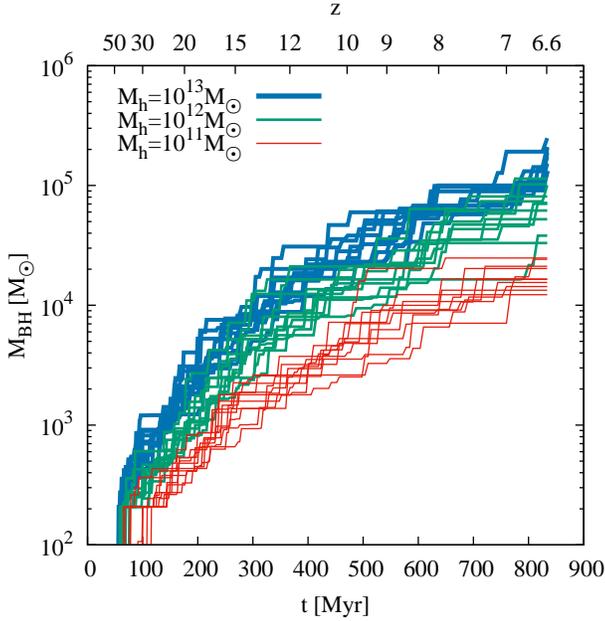} 
\caption{Growth history for Pop~III stellar remnant BHs that gain mass via galaxy mergers. Here we show 10 random realisations for each halo mass of $10^{11}\Msun$ (red), $10^{12}\Msun$ (green), and $10^{13}\Msun$ (blue). The BHs can reach final masses of $10^4-10^5\Msun$ by $z=6.6$.}
\label{fig:zMBH}
\end{figure}

While an accreting BH can also explain the line luminosities from CR7, these observations cannot yet differentiate between seed BH formation mechanisms. We first study BH formation in the progenitor haloes of CR7 by mergers of Pop~III stellar remnants. We follow the mass assembly history of these BHs in merger tree realisations and illustrate 30 randomly selected histories for three host halo masses in Figure \ref{fig:zMBH}.
The Pop~III remnant BHs grow to $10^4-10^5\Msun$ by $z=6.6$, depending on the mass of the halo. Since we only account for mass growth due to mergers of BHs, these values should be treated as an lower limit. We discuss the effect of additional mass accretion in Section \ref{sec:cav}. For the $10^{12}\Msun$ halo, we expect a BH with a mass of $\sim 10^5\Msun$ at $z \approx 7$, which could explain CR7 \citep{pallottini15,pacucci16}. Note that this quantifies the most massive BH in all progenitors of CR7. In Section \ref{sec:comp} we investigate explicitly those BHs that reside in clump A. Previous studies assume that the BH can accrete low metallicity gas down to $z=6.6$. This is a strong assumption, and we show in Section \ref{sec:metalfree} that the host haloes of stellar remnant BHs of this mass are generally polluted before $z=6.6$. 

\subsection{Direct collapse black hole}

\begin{figure}
\centering
\includegraphics[width=0.47\textwidth]{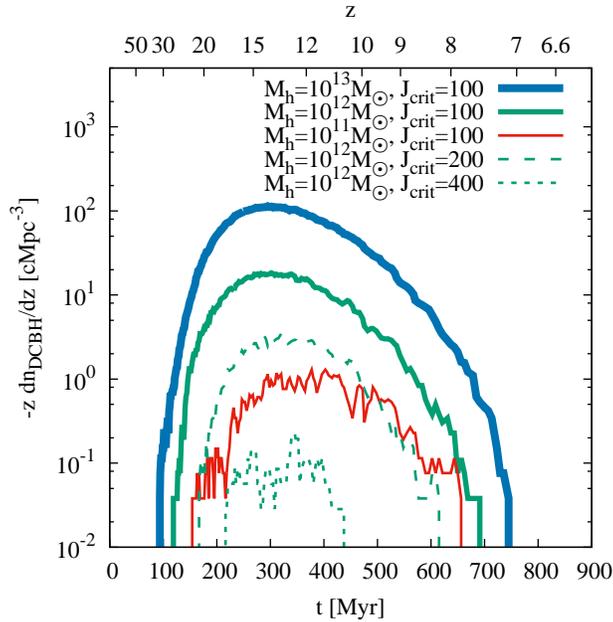}
\caption{DCBH formation rate as a function of redshift for several halo masses and values of $J_\mathrm{crit}$. The formation rate is normalised to the comoving volume of the main halo, where the red, green, and blue lines are for halo masses of $10^{11}\Msun$, $10^{12}\Msun$, and $10^{13}\Msun$, respectively. The solid, dashed, and dotted green lines show the numbers of DCBHs per redshift for $J_\mathrm{crit}=100,200,400$, respectively. The comoving volume is calculated as the mass of the main halo $M_h$ divided by the average cosmic density. In all these cases, we can form DCBHs during the assembly of the halo. A larger halo mass and a lower value of $J_\mathrm{crit}$ facilitates the formation of a DCBH down to lower redshifts. In none of the cases can we form a DCBH at redshifts $z<7.3$.}
\label{fig:zdndz}
\end{figure}

We also consider under which conditions a DCBH can form in the progenitor haloes of CR7. The DCBH formation rate density is shown in Figure \ref{fig:zdndz}.
DCBHs form in the five models at redshifts $z\geq 7.3$. Generally, a higher halo mass and a lower value of $J_\mathrm{crit}$ facilitates the formation of DCBHs. This is because a higher halo mass yields more progenitors and hence more possible formation sites for a DCBH, whereas for a lower value of $J_\mathrm{crit}$ haloes with lower masses can provide sufficient flux to enable isothermal collapse. The peak formation rate of DCBHs is around $z=15$ and the rate decreases steeply at $z<10$. In our models no DCBH can form at $J_\mathrm{crit}>600$ in the progenitors of CR7, whereas this is also limited by the finite number of merger tree realisations. Our implementation yields results similar to those by \citet{dijkstra14} and \citet{habouzit15}. The difference emerges from our implementation of Pop~III star formation and the fixed time of $10$\,Myr that we require for an atomic cooling halo to collapse, whereas they assume a redshift-dependent collapse time, which is generally longer. Hence, our assumptions are more optimistic, and similar to \citet{agarwal12,agarwal14}.

The difference between our results and those of \citet{agarwal15c} arise mainly from the treatment of metal enrichment. They assume a constant velocity for the enriched winds that yields a window of 50\,Myr in which DCBH formation is possible before the line-cooled halo is polluted with metals. In our model, we follow the pollution of individual haloes self-consistently. However, the important question is not only if DCBHs form in the progenitor haloes of CR7, but also if those haloes can remain below $Z_\mathrm{limit}$ for long enough. Otherwise, we should see the imprint of those metals in the spectrum.

\subsection{Mass of metal-poor gas}
\label{sec:metalfree}

\begin{figure}
\centering
\includegraphics[width=0.47\textwidth]{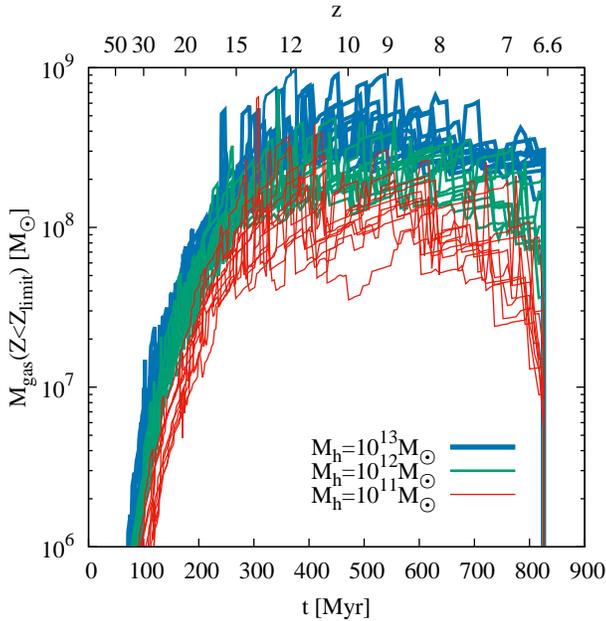} 
\caption{Maximum gas mass in the given halo with $Z<Z_\mathrm{limit}$. We plot 10 random realisations for final halo masses of $10^{11}\Msun$ (red), $10^{12}\Msun$ (green), and $10^{13}\Msun$ (blue). Even at low redshifts, it is possible for haloes to have a low metallicity and a gas mass of $\sim 10^8\Msun$.}
\label{fig:zMpristmax}
\end{figure}

\begin{figure}
\centering
\includegraphics[width=0.47\textwidth]{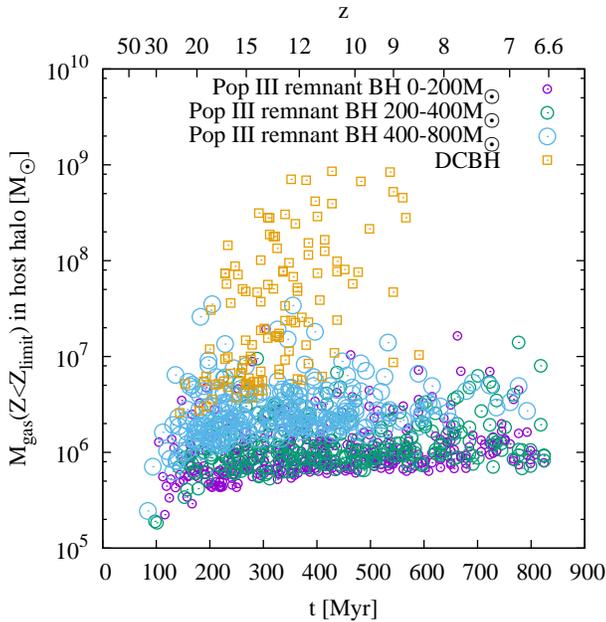}
\caption{Mass of metal-poor gas ($Z<Z_\mathrm{limit}$) surrounding BHs of different origin. The squares denote DCBHs, the circles are Pop~III stellar remnant BHs and the colours and sizes of the circles indicates the mass of the BH. The data is taken one time step before the haloes are polluted with metals to $Z>Z_\mathrm{limit}$, so the symbols mark out to which time a given type of BH can be surrounded by the indicated mass of metal-poor gas. For the red and green circles we plot only every 50th data point. By construction, the host halo of a DCBH is metal-free at the moment of formation, but shortly afterwards it generally merges with the enriched halo that previously provided the required LW flux.}
\label{fig:zMgasBH}
\end{figure}

We have shown in the previous sections that DCBHs and Pop~III remnant BHs can reach the masses needed to explain the \Lya and the \HeII luminosities of CR7. However, another important constraint are the upper limits on the O{\sc iii}]$1663\mathrm{\AA}$ and C{\sc iii}]$1908\mathrm{\AA}$ lines. The absence of these recombination lines requires the photon source to be embedded in metal-poor gas at $z=6.6$. In this section, we quantify the amount of low-metallicity gas that is present in the progenitors of CR7. We use the fiducial model and account for gas in haloes with a metallicity $Z<Z_\mathrm{limit}$. In previous sections we analysed the total mass of low-metallicity gas at a given redshift in all progenitors, but now we examine the maximum mass of metal-poor gas in a single halo at a given time. Its evolution over time is shown in Figure \ref{fig:zMpristmax}.
This plot shows that metal-poor haloes with a gas mass of $\sim 10^8\Msun$ are present down to redshift $z=6.6$. The more important question is if these metal-poor haloes can also host a sufficiently massive BH to explain the observed line luminosities. To answer this question we show the mass of low-metallicity gas in haloes that host either a DCBH or a Pop~III BH remnant in Figure \ref{fig:zMgasBH}.
Here, we plot the mass of low-metallicity gas surrounding BHs at the moment before the halo is polluted to $Z>Z_\mathrm{limit}$. This enables us to quantify the lowest redshift at which a BH can reside in a pocket of low-metallicity gas. For the Pop~III remnant scenario, we expect BHs in haloes with a gas mass of $\lesssim 10^7\Msun$ down to $z = 6.6$ but these BHs have masses of $\lesssim 800\Msun$, much smaller than the required mass of $\geq 10^5\Msun$. This is mainly due to the fact that either the BH progenitor itself pollutes its host halo with metals at the end of its lifetime or that another Pop~III star in the same halo enriches it with metals. The only possibility to remain below $Z_\mathrm{limit}$ is to have Pop~III stars that collapse directly to a BH. In our model, Pop~III BH remnants only grow by mergers with other BHs and might be polluted during the merger. Only a few BHs merge and remain in a low-metallicity environment. We find few Pop~III remnant BHs that can grow to $\sim 800\Msun$ before their host halo is polluted to $Z > Z_\mathrm{limit}$.

In contrast, the DCBHs reside by construction in low-metallicity haloes with a much higher gas mass of up to $\sim 10^9\Msun$. These higher masses can be reached because we quench Pop~III star formation in the host halo with a strong LW flux from a nearby star-forming halo. We find that DCBHs reside in a metal-free environment for only several tens of Myr before they are polluted by merging with the neighbour halo. DCBHs therefore reside in low-metallicity haloes only down to $z\approx 8.5$, which is $\sim 240$\,Myr before $z=6.6$. When interpreting this result, one should keep in mind that the rapidly-changing metallicity of the host halo due to the merger is an artefact of our recipe for DCBH formation. For this analysis, we use an optimistic value of $J_\mathrm{crit}=100$. The actual value can be up to an order of magnitude higher (Section \ref{sec:cav}) so the derived values should be treated with caution, since a higher $J_\mathrm{crit}$ might inhibit the formation of DCBHs.

\subsection{Comparison of scenarios}
\label{sec:comp}
\begin{figure}
\centering
\includegraphics[width=0.47\textwidth]{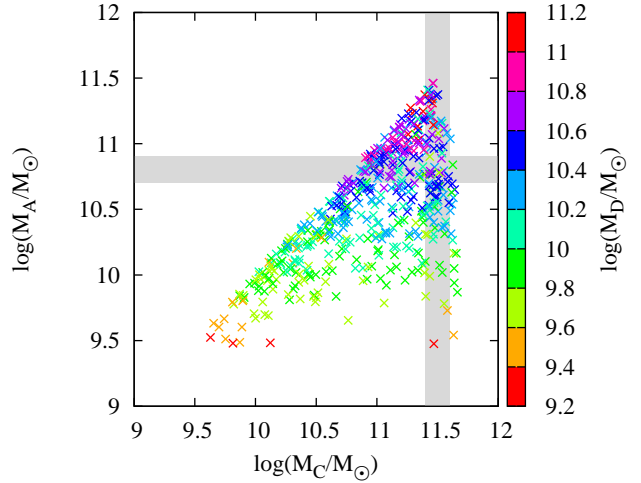}
\caption{Masses of the second, third, and forth most massive haloes at $z=6.6$, assuming that they merge to one halo within $100$\,Myr. The grey region indicates the required mass range for clump A and C of CR7 and we show 512 different merger tree realisations. Since we do not see a fourth clump within a projected distance of $\sim 10$\,kpc to CR7, we require that the fourth most massive clump in the merger tree is significantly less massive than clump A.
In this sample, we find constellations in the right mass range and with $M_D/M_A \lesssim 0.1$. The most massive clump B has always a mass of $\sim 10^{12}\Msun$.}
\label{fig:comp}
\end{figure}

\begin{figure}
\centering
\includegraphics[width=0.47\textwidth]{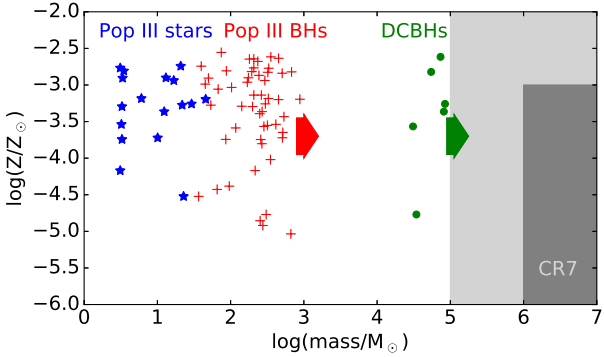}
\caption{Comparison of the three different models, regarding the metallicity of clump A at $z=6.6$ and the mass of the stellar population or of the BH, respectively.
The grey shaded area indicates the region in which we expect the system to have the same observational signature as CR7. A Pop~III stellar population has to have a mass of $\sim10^7\Msun$ and a BH should be in the mass range $10^5-10^7\Msun$ to yield the observed line luminosities. The upper limit of the metallicity is set by $Z_\mathrm{limit}$, in order not to violate the observed metal tax, but lower values are favourable. BHs of different origin can reside in sufficiently metal-poor halos, but, without including growth by accretion neither Pop~III remnants nor DCBH are sufficiently massive to account for the observations. The arrows for the BH populations symbolise possible mass growth by gas accretion; DCBHs require less mass growth by accretion to account for the spectral properties of CR7.}
\label{fig:sum}
\end{figure}

In this section, we compare the three models and estimate, which have the highest probability to reproduce the observations of CR7. To do so, we assume that CR7 is an ongoing merger of the three clumps A, B, and C, which merge at $z_0=6.0$. This is the most plausible scenario to obtain the right constellation of masses in CR7 as shown in Figure \ref{fig:comp}.
The most massive halo B always fulfils the requirement of $M_B \approx 10^{12}\Msun$, whereas only $<10\%$ of the merger tree realisations yield masses in the right range for halo A and C. As an additional constraint, we require a significant gap between the masses of the third and the fourth most massive halo at $z=6.6$, because we only see three clumps in CR7 and there is no evidence of a fourth equally massive clump. Hence, we assume that the fourth most massive halo D in the merger tree should be at least an order of magnitude less massive than halo A ($M_D/M_A \lesssim 0.1$).
This mass distribution is the anticipated constellation of CR7. We analyse the metallicity, and estimate the masses of both Pop~III stars (with the fiducial model) and BHs resulting from the two different seeding scenarios in halo A at $z=6.6$. The results are shown in Figure \ref{fig:sum}.
For a given scenario, this plot illustrates the possible mass-metallicity combinations of the three scenarios. For Pop~III stars, the stellar mass in clump A at $z=6.6$ remains always below $10^3\Msun$ and the halo has a metallicity of $\sim 10^{-3}$. Pop~III remnant BHs might reside in metal-poor halos at $z=6.6$, but are not massive enough ($\lesssim 10^3\Msun$) to explain CR7. Only the DCBH scenario can explain the observed line luminosities. For the given mass constellation of the halos A, B, C, and D, the probability that clump A hosts Pop~III stars is 37\%, that it hosts a Pop~III remnant BH is 98\% and that it hosts a DCBH is 10\%. These probabilities add up to over 100\%, because due to merging of the progenitor halos, clump A can contain Pop~III stars and BHs at the same time. Comparing these results to, e.g., Figure \ref{fig:zMBH} shows that it is crucial to treat CR7 as three individual clumps, since the BH mass in clump A is generally lower than the most massive BH in all CR7 progenitor halos. The third most massive halo A is less affected by metal enrichment, but the Pop~III remnant BHs do not merge to masses above $10^3 \Msun$.
Under these conditions and with the optimistic assumption of $J_\mathrm{crit}=100$ we find that $\sim 90\%$ of the DCBHs reside in sufficiently metal-poor gas at $z=6.6$.


Only $\sim 0.5\%$ of all the merger tree realisations, yield the right mass constellation and out of these, only $9\%$ produce results that are consistent with the observations of CR7. According to the Press-Schechter formalism, the number density of the host halo ($M_h=1.2 \times 10^{12}\Msun$ at $z=6.0$) is $n \lesssim 10^{-5}\,\mathrm{Mpc}^{-3}$. This yields an expected abundance of $n \lesssim 5 \times 10^{-9}\,\mathrm{Mpc}^{-3}$ for objects with the same constellation and observational signature as CR7. These estimates might even be lower with a smaller value of $Z_\mathrm{limit}$, suggested by the non-detection of C{\sc iii}]. The survey of \cite{matthee15} covers $5\,\mathrm{deg}^2$ in the redshift range $z=6.5-6.6$, which corresponds to an observed volume of $\sim 4.3 \times 10^6\,\mathrm{Mpc}^3$. The volume in the COSMOS fields, where CR7 was found,  is $\sim 1.5 \times 10^5\,\mathrm{Mpc}^3$ and the expected number density of CR7-like sources is of the order $10^{-6}-10^{-7}\,\mathrm{Mpc}^{-3}$ \citep{pallottini15,visbal16}.
CR7, therefore, appears to have been a fortunate discovery, for the design of that survey. Larger surveys can confirm whether the mechanisms suggested in this paper occur with the expected probability or if we might have to adopt the model and assumptions to account for such rare sources.

There are other galaxies near the epoch of reionisation that have comparable observational signatures. \citet{matthee15} find more than ten \Lya emitter candidates potentially similar to CR7, which are worth being further investigated. Also the `Himiko' galaxy  is very bright and extended in \Lya, consists of three clumps of which one is very blue, and shows no sign of metals emission lines in the rest-frame UV \citep{ouchi09,ouchi13,zabl15}. However, no \HeII emission has be confirmed for this galaxy, which disfavours Himiko as a potential host of Pop~III stars.

\section{Caveats}
\label{sec:cav}

The main shortcoming of our semianalytical approach is the lack of spatial information in the merger tree and hence the simplified treatment of metal enrichment. We check statistically if a halo is polluted, which yields reasonable results on average over many merger tree realisations. Once a halo is polluted we assume that the metals mix homogeneously with the gas and assign a single metallicity to the halo. However, metal-enriched winds may not mix effectively with the dense gas in minihaloes and a large fraction of this gas might remain at low metallicity \citep{cen08,pallottini14,ritter15,smith15}. Three-dimensional hydrodynamical simulations are required to obtain a more reliable metal enrichment history and identify pockets of metal-free gas.

We do not explicitly account for accretion onto BHs. In our model, the DCBHs have a seed mass of $10^4-10^5\Msun$ and the Pop~III remnant BHs only gain mass by mergers. The accretion of large masses of gas is necessary to account for the observed BH mass density at $z=6$ \citep{tanaka09}. Pop~III remnant BHs form in minihaloes with shallow potential wells, which are unable to retain the photoionised heated gas \citep{johnson07,alvarez09}. Also the energy of a single supernova from a massive star is sufficient to clear the halo of accretable gas. The average time that Pop~III remnant BHs can accrete mass in our model is $\sim 400$\,Myr. To grow from a seed mass of $\sim 200\Msun$ to a final mass of $10^6\Msun$ at $z=6.6$ they would have to accrete constantly at $80\%$ the Eddington rate. Using cosmological zoom-in simulations, \citet{jeon12} study the assembly of the first galaxies under feedback from a central BH. They find that the accretion rate on to Pop~III remnant BHs remains always below $10^{-6}\Msun\,\mathrm{yr}^{-1}$ for a radiative efficiency of $\epsilon = 0.1$. This is below the required accretion rate to account for the BH in CR7 at $z=6.6$. Hence, the mass of the Pop~III remnant BHs in our model might generally be higher, but they can not grow sufficiently massive to account for the observations \citep[but see][]{lupi16}.

In our analysis, we investigate $J_\mathrm{crit}=100-400$ as the LW flux required to enable isothermal direct collapse of an atomic cooling halo. The probability for forming a DCBH is a steep function of this value \citep{inayoshi14c} and we do not find any DCBHs in our model at $J_\mathrm{crit}>600$. The required flux varies from halo to halo \citep{shang10,latif14a} and also depends on the stellar spectrum. Pop~III stars with effective temperatures of $\sim10^5\,$K require $J_\mathrm{crit}\approx10^5$, whereas subsequent stellar populations have lower effective temperatures and require a smaller $J_\mathrm{crit}$ \citep{bromm03,shang10,sugimura14,agarwal15a,latif15a}. In our model, a young population of Pop~II stars provides the required flux for isothermal direct collapse. We therefore expect $J_\mathrm{crit}<1000$ \citep{shang10,agarwal15b}, but the exact value remains an open question.

A third way to form SMBHs in the early Universe is the fragmentation of gas in a low-metallicity halo into a dense nuclear cluster of low-mass stars and the subsequent build-up of a $\sim 10^3\Msun$ star via runaway mergers in the cluster \citep{portegies04,devecchi09,lupi14}. This scenario results in the creation of a $10^3\Msun$ BH and requires enough low-metallicity gas to form the cluster.  Although this channel can generally produce as many BHs as Pop~III BH remnants, it faces the same problem as the other scenarios: how can such a BH grow to sufficient masses without the host halo being polluted by metals? Given the formation criteria for a dense cluster we expect that it might yield results comparable to those of the Pop~III remnant scenario and hence not be able to explain the properties of CR7.

To create the merger trees, we use a code based on \citet{parkinson08}. This method is tuned to reproduce the halo mass function of the Millennium simulation \citep{springel05} of intermediate and high-mass halos at redshifts $z \lesssim 4$. For higher redshifts ($z\approx 30$), this method tends to yield slightly more minihaloes, which are possible formation sites of Pop~III stars. Using the original method by \citet{cole00}, which is not calibrated to match the halo mass function of the Millennium simulation, we form fewer primordial stars at $z>15$. This leads to a metal enrichment at later times and a higher SFR for Pop~III stars at $z \sim 10$. For $z<8$, the mass of pristine gas is roughly the same in both models so there is no difference in the probability of finding a massive primordial halo at $z=6.6$. The method of \citet{parkinson08} yields total stellar masses of Pop~III stars at $z=6.6$ that are about a factor of two lower compared to the original method by \citet{cole00}, but this is not sufficient to account for the lack of luminosity. The mass function of minihaloes at high redshift still remains an open question and better constraints might help to improve our model of Pop~III star formation.

\section{Conclusion}
\label{sec:conclusion}

We have explored the nature of CR7, a \Lya emitter at $z=6.6$ \citep{matthee15,sobral15}, which has very strong \Lya and \HeII signatures without any detection of metal lines in the spectrum. Using a semianalytical merger tree model, we have investigated a variety of formation histories for CR7 and tested different scenarios for its origin.

\citet{sobral15} originally proposed that a recent Pop III starburst with a total stellar mass of $\sim 10^7\Msun$ or an accreting BH can account for the observational constraints. Based on our current understanding of Pop~III star formation, we show that such a starburst seems not to be possible for several reasons. The mass of metal-free gas decreases with time since more and more haloes are polluted by SNe. Hence, after peaking around $z\sim 15$, the cosmic SFR for Pop~III stars declines sharply at lower redshifts. Moreover, very hot massive stars are required for the \HeII line emission and the short lifetimes of these stars consequently requires a very recent burst, $\lesssim 1$\,Myr to be in agreement with the observed EW of \HeII. Our model fails to reproduce the observed \HeII line luminosity by about 10 orders of magnitude. Besides this fiducial model of primordial star formation, we also tested various models in which Pop~III stars form in gas with a metallicity $Z<Z_\mathrm{crit}$, in which star formation is based on the merger history of the halo, and in which we adopt a different IMF. None of these models can explain CR7 as a primordial star cluster.

If CR7 hosts Pop~III stars, the metal pollution by these first stars must be significantly less efficient than previously assumed and we need a mechanism to form $10^7 \Msun$ of metal-free stars synchronised in one halo within $\sim 1$\,Myr. Recently, \citet{visbal16} propose that photoionisation feedback could prevent early star formation and hence enable the collapse of a $\sim 10^9 \Msun$ halo at $z \approx 7$.

We also investigate the possibility that CR7 hosts an accreting BH. This scenario seems more appealing than Pop~III stars because DCBHs form down to $z\sim 7.3$ and Pop~III BH remnants can form down to $z=6.6$. Several groups have shown that such BHs can reproduce the observational constraints \citep{pallottini15,agarwal15c,smith16,dijkstra16,smidt16,pacucci16}.  However, they assume that the BH is embedded in metal-poor gas at $z=6.6$, which is difficult to obtain with our model because when a Pop~III star BH forms, it is very likely that a SN also enriches the halo with metals. Furthermore, we assume that Pop~III star BHs mainly grow via mergers and consequently the already low possibility of finding a BH in a metal-poor halo shrinks with every merger with another halo, which could be enriched with metals. Finally, we only find Pop~III BH remnants with masses $<10^3\Msun$ in low-metallicity haloes with gas masses of $\sim10^6\Msun$ at $z=6.6$.

The most promising explanation for CR7 is an accreting DCBH. By construction, they form in a metal-free halos and may remain in metal-poor gas until $z=6.6$. Under the optimistic condition of $J_\mathrm{crit}=100$, we find that a small fraction of systems can host DCBHs that are able to reproduce the line luminosities of CR7 without violating the upper limit of the metal line luminosities. Our findings are supported by other observational constraints. Only an accreting BH can account for the spatial extension of the \Lya emitting region and for the velocity offset between the \Lya and \HeII line peaks, because this velocity offset requires a source lifetime of $>10$\,Myr \citep{smith16}. A stellar source in CR7 would however require a recent burst of $\lesssim 1$\,Myr, and a very low metallicity of $<10^{-7}$ to account for the large EW of \HeII \citep{schaerer03,raiter10}. We note that we are not able to confirm, if an accreting BH is able to produce the EW of \HeII, which is an important question to address in the future.


There are other observations that might help to better understand the nature of CR7 and distinguish between the different formation scenarios. The dwarf galaxy I~ZW~18 at a distance of $18$\,Mpc also shows strong \HeII emission and has a very low metallicity \citep{kehrig15a,kehrig16}. It might host either metal-free or Wolf-Rayet stars, which could account for these observational features. Smaller halos, such as I~ZW~18, might remain unpolluted by their progenitors down to lower redshift and e.g. photoionisation heating might prevent star formation for a long time \citep{visbal16}. The study of the assembly history of such systems might reveal interesting new insights that can help to understand more massive low-metallicity counterparts at higher redshift.

Another observational signature of whether CR7 hosts a BH or a Pop~III stellar population is the X-ray flux.
The expected luminosity from X-ray binaries is $L_\mathrm{x} \approx 10^{40}\,\mathrm{erg}\,\mathrm{s}^{-1}$ for a SFR of $1\Msun\,\mathrm{yr}^{-1}$ \citep{glover03,grimm03,mineo12}, which is at least two orders of magnitude lower than the emission from a BH of $\sim 10^6\Msun$. CR7 was not detected as a point source in the Chandra COSMOS Survey. Therefore its  X-ray luminosity in the energy range $0.2-10$\,keV is less than $10^{44}\,\mathrm{erg}\,\mathrm{s}^{-1}$ \citep{elvis09}. For a BH with a mass of $10^6\Msun$, accreting at $40\%$ of the Eddington rate, and a bolometric correction based on \cite{marconi04} we find $L_\mathrm{x} \lesssim 2 \times 10^{42}\,\mathrm{erg}\,\mathrm{s}^{-1}$ in the observer rest frame $0.2-10$\,keV band. This translates into a flux of $F_\mathrm{x} \lesssim 8.2 \times 10^{-18}\,\mathrm{erg}\,\mathrm{cm}^{-2}\,\mathrm{s}^{-1}$ or $\sim 1.4 \times 10^{-7}$ counts per second with Chandra in the $0.2-10$\,keV band. Hence, integration times of several years might be required, which seems infeasible until the next generation of X-ray telescopes, such as {\sc Athena} \citep{athena13}, could approach this problem with a significantly higher effective collecting area. Alternatively, if many more DCBH candidates similar to CR7 are found, one could stack deep X-ray observations of these objects and attempt to detect their X-ray emission using the stacked image. In this case, a successful detection would imply that some significant fraction of the candidates are indeed DCBHs, although it would not allow us to say with certainty that any particular candidate is a DCBH. With this method, however, there is no X-ray detected in the stack of Ly$\alpha$ emitters in the Chandra COSMOS Legacy survey \citep{civano16}, down to a total depth of $1.38$\,Ms corresponding to a luminosity of $10^{43}\,\mathrm{erg}\,\mathrm{s}^{-1}$ (at the redshift of the sources) in the $0.5-2$\,keV band (Civano, private communication). Another method to identify DCBHs was recently proposed by \citet{pacucci16}. They identify two objects with a robust X-ray detection found in the CANDELS/GOODS-S survey with a steep spectral slope in the infrared and an extremely red spectra. They argue, based on their assumptions, that this can be explained by either a DCBH or an atypically high SFR of $> 1000 \Msun \, \mathrm{yr}^{-1}$. Another explanation of this spectra could be a dusty AGN without the constraint of being a DCBH.

More crucially, validation of the general scenario of a low metallicity in CR7 requires further deep spectroscopy to estimate the level of metallicity and the source of the hard spectrum in CR7. The focus should hence be to obtain deeper spectra from UV to near-IR to improve the limits on the metal line emission and to further constrain the different formation histories.
Beyond current instruments, such as MOSFIRE and X-SHOOTER, NIRSpec on JWST will be the ideal instrument to probe sources such as CR7, which provide a tantalizing preview into the initial episodes of star and BH formation, which will fully be elucidated in the near future with an array of next-generation observational facilities.



\subsection*{Acknowledgements}

The authors would like to thank Julia Gutkin, Aida Wofford, Anna Feltre, Dan Stark, Jorryt Matthee, M\'elanie Habouzit, Andrea Negri, Carolina Kehrig, and Francesca Civano for valuable discussions and helpful contributions. The authors acknowledge funding from the European Research Council under the European Community's Seventh Framework Programme (FP7/2007-2013) via the ERC Grant `BLACK' under the project number 614199 (TH, MV) and via the ERC Advanced Grant `STARLIGHT: Formation of the First Stars' under the project number 339177 (MM, DJW, RSK, SCOG, EWP). SCOG, RSK, and EWP acknowledge support from the DFG via SFB 881, `The Milky Way System' (sub-projects B1, B2 and B8) and from SPP 1573 `Physics of the Interstellar Medium'. MAL has received funding from the European Union's Horizon 2020 research and innovation programme under the Marie Sklodowska-Curie grant agreement No. 656428. VB was supported by NSF grant AST-1413501.

\bibliographystyle{mn2e}

\bibliography{CR7}

\bsp
\label{lastpage}

\end{document}